\documentclass[%
 preprint, linenumbers,
 amsmath,amssymb,
 aps, physrev,
]{revtex4-2}

\usepackage{graphicx}
\graphicspath{{images/}}
\usepackage{dcolumn}
\usepackage{bm}
\usepackage{amsthm}
\usepackage[autostyle]{csquotes}
\usepackage{caption}
\usepackage[labelformat=simple]{subcaption}   
\newtheorem{theorem}{Theorem}[section]

\newtheorem{lemma}[theorem]{Lemma}

\newtheorem{definition}[theorem]{Definition}

\begin{document}

\preprint{APS/123-QED}

\title{Representational drift changes the encoding of fast and slow-varying natural scene features differently}
\author{Siwei Wang}
\affiliation{Department of Neurobiology and Behavior, SUNY Stony Brook University \\ National Institute for Theory and Mathematics in Biology, Northwestern University and The University of Chicago}
\author{Elizabeth A.\ de Laittre}
\affiliation{Committee on Computational Neuroscience, University of Chicago}
\author{Jason N.\ MacLean}
\affiliation{Committee on Computational Neuroscience, University of Chicago\\Department of Neurobiology, University of Chicago \\ National Institute for Theory and Mathematics in Biology, Northwestern University and The University of Chicago}
\author{Stephanie E.\ Palmer}
\affiliation{Department of Organismal Biology and Anatomy, University of Chicago\\ Department of Physics, University of Chicago \\Physics Frontier Center for Living Systems, University of Chicago \\ National Institute for Theory and Mathematics in Biology, Northwestern University and The University of Chicago}

\date{\today}
\nolinenumbers 
\begin{abstract}
Representational drift refers to an unstable mapping between neural activity and input sensory or output behavioral variables. While much work has focused on the effect of representational drift on single, simple external variables, we investigate the differences in representational drift across spatiotemporal features in a moving visual stimulus. The neural responses across animals to the same movie reflect both common, encoded stimulus features and idiosyncratic individual variation. To extract the shared neural encoding of stimulus features only, we learn a latent space embedding using weakly supervised contrastive learning. This approach pulls neural activity together in the embedding space if they are responses to the same stimulus segment and push them apart if not. This approach enables us to probe how stimulus features fluctuating as fast as 33 ms (the movie frame rate) are encoded by variable neural codes across animals. It also allows us to investigate how representational drift changes the encoding in individuals across sessions. We observe that our learned embedding is near-optimal for decoding natural features (background scenery, local motion, complex spatio-temporal features, and time) and neural activity from novel animals. This suggests that our embedding retains the encoding of multiple features at higher temporal granularity compared to previous methods. To quantify representational drift, we apply the trained decoder (which achieves near-optimal performance in one session) to a subsequent session recorded 90 minutes later. We then use the decrease in decoding performance as a proxy for the magnitude of drift. We show that the drift changes the encoding of fast-varying local motion features at a rate 5-6 times higher than slower-varying scenery features. Drift also perturbs the local geometry in the embedding. By exploring drift across multiple features that change over different timescales, we provide initial evidence that the rate of drift varies with the temporal statistics of scene features.

\end{abstract}

\maketitle

\section*{Introduction}

Representational drift refers to a shifting relationship between external variables and neural activity observed on timescales ranging from minutes to weeks \citep{Deitch2021}. It has been documented across a wide range of sensory and motor systems. In the hippocampus, drift manifests as gradual changes in neural firing patterns that provide a temporal context for events, supporting the refinement of memory for temporal order \citep{Manns2007}. In sensorimotor and motor areas, drift has been interpreted as a reflection of circuit malleability that enables continual learning despite stable behavior \citep{Rule2019, Driscoll2022, Natrajan2025}. These findings have motivated theoretical efforts to identify biologically plausible mechanisms of drift. On timescales of days to weeks, proposed mechanisms include neuronal or synaptic turnover (e.g., the integration of new neurons or synaptic rewiring \citep{Devalle2025, Driscoll2022}), homeostatic and plasticity-driven circuit reorganization \citep{Micou2023}, and slow fluctuations in intrinsic excitability \citep{Morales2025}. Although such models offer insight into why drift occurs over particular timescales, they leave open the question of whether drift has a homogeneous or heterogeneous effect on the neural code across timescales. For example, recent theoretical work \citep{Morales2025} suggests that a fast Hebbian learning rule may contribute to drift in the early phase of learning (minutes/hours) while stabilizing representations in the long term (days/weeks). Despite progress, there is no consensus on whether representational drift reflects stochastic noise in neural coding (a bug), whether it reshapes neural representations towards better fitness throughout evolution (a feature), or if drift at different timescales serve different functional roles. 

Representational drift in sensory systems presents a distinct readout challenge. Unlike in memory or motor circuits, where drift often preserves task-relevant information despite changes in neural activity \citep{Driscoll2017, Rule2020, Aitken2022} (though notably, single-day decoders can fail to generalize across days, c.f.\ \citep{Rule2020}), drift in sensory representations directly impairs the encoding of external variables. For example, in the piriform cortex, odor-evoked responses drift over days such that the performance of a fixed linear odor decoder trained on an initial session drops to chance levels after approximately one month \citep{Schoonover2021}. In the primary visual cortex (V1), representational drift has been shown to depend on stimulus statistics: responses to artificial stimuli remain stable, whereas responses to naturalistic stimuli exhibit substantial drift \citep{Marks2021}. This suggests that stability in V1 is not an inherent property of individual neurons, but depends on the complexity of its inputs. Several studies have explored whether stable representations of external variables exist in V1 activity despite drift. \citet{Xia2021} reported that when V1 population activity is binned into 1-second windows, a fixed linear decoder could reliably identify the corresponding segment of a repeated movie, even after weeks of drift. They interpreted this as evidence that V1 encodes a stable temporal sequence by embedding stimulus information in a low-dimensional manifold, with drift confined to dimensions orthogonal to stimulus encoding. However, this interpretation was later challenged by \citet{Sadeh2022}, who showed that neural activity is dominated by internal state fluctuations, such as arousal and movement, rather than external sensory inputs. Using large-scale recordings from the Allen Brain Observatory, they demonstrated that variability in pupil size, running speed, and other behavioral metrics is strongly correlated with the observed drift in V1 responses. This suggests that drift may primarily reflect changes in the internal state rather than changes in the encoding of external variables. 

However, to understand how representational drift affects the encoding of external visual features that are informative for visually-guided behaviors, we need to probe the neural code on faster timescales. Visual motion in natural scenes often evolves on timescales shorter than one second \citep{Wainwright2001, Wainwright2000, Salisbury2025}. Previous work also suggests that different visual features are relevant for different behaviors that operate at distinct timescales. For example, fine textural details typically can be accurately discriminated in tens to a few hundred milliseconds \citep{Resulaj2018}. In mice, V1 takes only $\sim$80~ms to start the process of texture or pattern discrimination, with behavioral performance saturating around 300~ms \citep{Resulaj2018}. In contrast, motion cues associated with sudden threats, such as looming stimuli or rapid optic flow, can evoke escape or freezing responses within just a few and tens of milliseconds. Mice have been shown to initiate defensive behaviors within $\sim$250~ms of looming stimulus onset \citep{Yilmaz2013}, and even transient expansive stimuli as brief as 1.6~ms can trigger reliable escape responses \citep{Zhang2014, Liang2015}. These findings indicate that the ethologically relevant timescales for visual processing spans at least two orders of magnitude: from 1--10~ms (for predator or prey detection) to several hundred milliseconds (for detailed scene analysis). Features at either end of this range (which is entirely sub-second) support distinct behavioral functions.

Identifying meaningful visual features from natural scenes is challenging due to their complexity. Both textural and motion feature distributions in natural environments are non-Gaussian and heavy-tailed \citep{Salisbury2025, Schwartz2001, Olshausen1996, Ruderman1994,Saremi2013}. Moreover, features at different spatiotemporal scales often interact. For example, the local motion of an object is embedded within the broader context of a scene, and changes in viewpoint or occlusion can couple these features. This complexity makes it intractable to manually enumerate or label all possible feature states, even within a short movie sequence (e.g. 100 frames or 3.3 seconds at a 30 Hz frame rate). To overcome this challenge, we apply a task-agnostic, representation learning approach that does not rely on predefined feature labels. Specifically, we use cross-modality contrastive learning \citep{tian2020contrastive, CLIP2021} to construct a shared embedding space between neural activity and the corresponding segments of a movie. Contrastive learning identifies shared structure in different representations of the same underlying event. A canonical example is OpenAI's CLIP model \citep{CLIP2021}, which learns to associate an image (e.g., a photo of a dog) with its textual description (\enquote{a dog on bench}) in a shared embedding space, without requiring class labels. Analogously, we train an embedding by pairing population activity from V1 with aligned movie segments. The key idea is to define similarity through temporal co-occurrence. Neural activity and movie segments from the same time window are pulled together as matching pairs, while mismatched pairs (from distant time points) are pushed apart. In other words, the \enquote{togetherness} between neural activity in response to a specific movie segment defines the natural features they selectively encode. To optimize our framework for retaining components that correspond to external stimulus features, we also perform contrastive learning across animals. The learned embedding pulls neural activity from multiple mice viewing the same movie segment together as matching tuples. This cross-animal training constrains the learned embedding to emphasize the encoding of external stimulus features common across subjects, while de-emphasizing individual variability due to internal states such as pupil size or locomotor activity.

Using population recordings from mouse V1 from the Allen Brain Observatory, we demonstrate that our learned embedding achieves near-optimal performance ($\sim$99\% accuracy) in discriminating neural responses spaced only 33~ms apart, i.e.\ at single-frame resolution of the movie stimulus. The movie used in this dataset (a clip from the Orson Welles film \textit{Touch of Evil}) contains visual dynamics unfolding on multiple timescales. We observe that there are 69 local motion features (e.g., moving human/objects) that fluctuate rapidly (every 33--66~ms), while 42 scenery features (e.g., background textures that only change as the camera pans, see examples in Figure \ref{fig:sceneryexample}) vary more slowly (on the order of 500--1000~ms). These two features jointly span the ethologically relevant temporal window described above (there are 198 unique combinations of scenery$+$ motion features). Our previous work analyzing retinal encoding in a broad set of natural scenes showed that the combination of local motion and scenery features accounts for approximately 80\% of the entropy in decodable stimulus information \citep{wang2022learning}. We therefore treat these features and their interaction as a first-order proxy for the high-dimensional structure present in natural movies, including this Hollywood film. Our learned embedding achieves near-optimal decoding of these complex features. The learned representation also generalizes well across animals and trials. A decoder trained on one set of animals can accurately read out features from held-out animals and even from single-trial neural responses. 
Because our learned embedding can decode natural features nearly optimally, it gives us a way to quantify the magnitude of representational drift in natural stimulus scenes. Conventional single-neuron analyzes quantify the magnitude of representational drift through tuning curve changes \citep{Marks2021}, but this framework is not feasible for naturalistic stimuli where tuning curves are not well-defined. Given the near-optimal decoding performance, our learned embedding mitigates this limitation because any representational drift must \textbf{necessarily} result in a decrease in decoding performance. Applying linear decoders trained on the embeddings of session 1 to the data of session 2 (90 minutes later), we observed consistent performance decrease $\sim$ 50\% in all natural features. How representational drift changes neural coding exhibited striking feature dependence. The performance decrease of local motion features that fluctuate at 33-100 ms is 5-6 times worse than the scenery features. This finding extends previous work \citep{Rule2020, Driscoll2017, Schoonover2021, Deitch2021, Xia2021, Sadeh2022} and reveals the heterogeneous nature of representational drift with respect to different features.

We further show that without drift, the geometry of the learned embedding captures the frame-to-frame transition of spatiotemporal features throughout the movie. With drift, this geometry is disrupted. In particular, drift disrupts the geometry of features that fluctuate on the timescale of 1-3 frames (33-100 ms), which is why the performance of local motion features is dramatically decreased. Given the difference in dynamic timescales between local motion and scenery features, our results suggest that downstream systems may need segregated pathways or develop distinct compensation mechanisms to mitigate representational drift. Therefore, our work enriches the previous notion that the primary visual cortex has a multidimensional code for the external world. We further suggest that representational drift may act as an active sculpting force that encourages modular neural processing for sensory features of distinct spatiotemporal scales, rather than passive noise \citep{Driscoll2017, Rule2020, Sadeh2022}.

\section*{Results}
We analyze publicly available Neuropixels recordings from the mouse primary visual cortex (V1) during movie presentation, available from the Allen Brain Observatory visual coding dataset. The dataset contains two recording sessions separated by a 90-minute interval. We chose this dataset over calcium imaging data available for the same animals because, in the spiking dataset we analyze, the time interval between the two sessions is kept uniform at 90 minutes for all animals. This allows us to rule out the influence of individual bias resulting from behavioral variability and to focus on how the encoding of stimulus features changes between sessions (see Results. \ref{sec:movie} and \ref{sec:decode}). In addition, the Neuropixels dataset contains sessions with 30 repeats of the movie stimulus while there are only 10 repeats in the calcium imaging dataset. The higher number of repeats averages out trial-to-trial variability and allows us to resolve the direction of drift in feature space. In this Neuropixels dataset, we selected the primary visual cortex (V1) for our analysis because it has been sampled the most comprehensively \citep{Siegle2021, Allen2019, Deitch2021} (see Supplementary Materials \ref{method:pseudo} for details). V1 also has a dual role in visual processing. It transmits behaviorally relevant features to higher visual areas while also directly connecting to the subcortical structures that drive rapid survival behaviors \citep{Zhao2014, Liang2015}. We first estimate the range of spatiotemporal scales of stimulus features present in the movie. We then investigate how representational drift changes the encoding of these features across this range of spatiotemporal scales.

\subsection{The visual stimulus contains features with multiple spatiotemporal scales}
\label{sec:movie}

\begin{figure}
    \centering
    \includegraphics[scale=0.45]{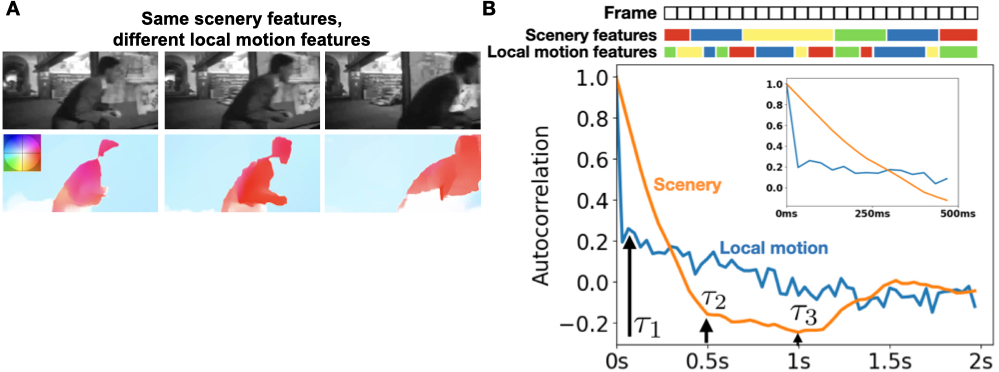}
    \caption{\textbf{Local motion and scenery features fluctuate at different timescales.}
   A) When we cluster static scenery frames and local motion frames separately, we find that frames with similar scenery content (sharing the same clustering label) often contain different local motion content (as determined by clustering procedures independently applied to scenery and local motion frames; see main text and Supplementary Materials~\ref{supp:clustering}). The three example static frames (top) all share the same scenery label, yet their corresponding local motion frames at time $t$ (bottom, they are calculated as the difference between frame $t$ and $t+1$) each have a different local motion label.  B) Autocorrelation decays differently for scenery and local motion labels (schematized at the top, with actual data below). Local motion features decay rapidly, reaching near-zero correlation after just 33-100 ms (1-3 frames). In contrast, scenery features show a much slower decay, reaching zero only after 400-500 ms (12-15 frames) and a negative peak around 1s (likely corresponding to scene changes in the movie). This analysis uses the first 400 frames (first half of the movie), with similar decay timescale patterns observed in the second half of the movie (see Supplementary Materials~\ref{autocorr2}).}
    \label{fig:movie}
\end{figure}

To overcome the computational challenge of parameterizing complex movies, we decomposed the visual stimulus into two streams: static scenery features and dynamic local motion features (Fig.~\ref{fig:movie}A). We derived the local motion features by calculating the pixel-by-pixel difference between adjacent movie frames, also known as `optic flow' (see Supplementary Materials~\ref{supp:clustering} for details). We then applied unsupervised hierarchical clustering \citep{JMLR:v12:pedregosa11a} independently to both streams. Clustering provides an estimate of how many discrete scenery and local motion features are present in the movie. In our subsequent analyzes, we use these cluster labels as a proxy for scenery and local motion features. In addition, clustering movie frames by scenery versus by local motion features yields different clusters, which enables us to construct three distributions of stimulus features in the movie stimulus. One uses clustering based on scenery features, another uses local motion features, and a third combines both sets of features. We choose clustering thresholds on both scenery and local motion features to ensure that 1) the entropy of the joint feature approximates the full entropy of time (the entropy of combined scenery+local motion features is 7.2 bits, 84\% of the full entropy for time $1/400\cdot \log_2(1/400) = $ 8.6 bits) and 2) the entropy of scenery and local motion features is comparable (The entropy for flow features is 5.7 bits, while the entropy for scenery features is 5.2 bits). As shown in Fig 1A (and scenery feature examples in \ref{fig:sceneryexample}), our clustering is also interpretable. Our local motion features correspond to local motion in different directions (shown in different colors). 

Comparing the autocorrelation timescales between streams reveals that the local motion features fluctuate approximately ten times faster than the scenery features (Fig.\ \ref{fig:movie})). Specifically, local motion features have an autocorrelation timescale of just 33--100 ms (equivalent to 1-3 frames in the movie). Because of this, we use $\tau_1$ = 66 ms as the characteristic timescale for local motion features. Scenery features fluctuate at much slower rates of 500 ms to 1000 ms. We define two timescales for scenery features, $\tau_2$ = 500 ms and $\tau_3$ = 1000 ms. This stark difference in autocorrelation time reflects the rapid changes in local motion (such as leftward or rightward movement of actors in the scene) compared to the relatively infrequent transitions between ``scenes'' (while the 30-second clip is one continuous camera shot with no cuts, it contains multiple distinct backgrounds). The short (33--100 ms) autocorrelation timescale of local motion features implies that adjacent frames often contain different local motion features, requiring analysis at single-frame resolution. Consistent with this, clustering the movie stimulus using coarse, 1-second windows, as in previous studies \citep{Xia2021, Sadeh2022, schneider2023cebra}, fails to capture rapid visual motion dynamics (see Supplementary Materials~\ref{autocorr2}). This motivates a frame-resolved analysis at 33 ms, which allows us to track how neural representations encode both fast local motion features and slower-varying scenery features.

\subsection{Learning a generalizable representation from weakly supervised, cross-modality contrastive training}
\label{sec:decode}
\begin{figure}
    \centering
    \includegraphics[scale=0.4]{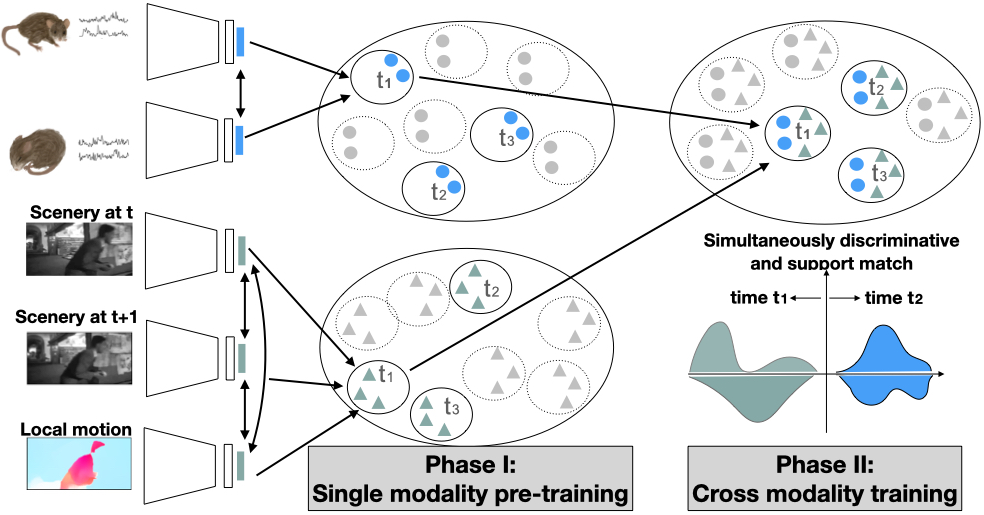}
\caption{\textbf{Weakly supervised contrastive learning extracts features from the movie and neural code.}. Our method uses temporal co-occurrence to bring together neural and visual representations from the same temporal window while separating those from different windows. The approach consists of two phases: First, single-modality training in which separate ResNet50 networks independently learn embeddings for neural activity and the visual stimulus. Second, cross-modality training that aligns the output embeddings between these ResNet50 networks. The goal here is to refine the learned embeddings so that samples from any modality that correspond to the same time bin are pulled close together and samples that occur at different time bins are pushed apart (each cluster of blue or green dots represents samples of different modalities from the same time, are pulled together). As a result, it produces a new, modified embedding with shared decision boundaries for stimulus features (e.g., times $t_1$ and $t_2$ shown in green and blue) while matching their supports across modalities (lower right panel). This support match ensures that the stimulus features in the embedding learned from neural activity precisely correspond to those in the visual stimulus and vice versa. The alignment enables assessment of the encoding of complex natural features without explicit parameterization. See Supplementary Materials \ref{alltoall}.}
\label{fig:intro}
\end{figure}

To analyze complex stimulus features without requiring explicit pixel-level parameterization, we developed a weakly supervised contrastive learning approach guided by temporal co-occurrence \citep{Chen2020b, CLIP2021}. Our method is inspired by an intuition/observation: samples of neural activity recorded at the same time in the movie but on different trials should encode the same stimulus features across trials, despite slight differences in the responses from individual animals. We implemented this principle through a two-phase training approach (Fig.\ \ref{fig:intro}). In the first phase, we applied contrastive learning to neural activity samples from different animals, bringing representations within the same temporal window together in the embedding space. This process extracts shared stimulus features while eliminating individual animal biases and variance \citep{Chen2020}. At the same time, we trained a separate model to efficiently represent the movie by contrasting its scenery and local motion features. Contrasting makes the learned embedding pull together scenery features and local motion features from the same sampling windows (centered at the same frame) while pushing apart those from different sampling windows. This was demonstrated in \citep{Tian2019} to obtain an efficient representation of the movie itself. This single-modality phase develops separate and efficient representations for both neural activity and stimulus features in the movie. In the second phase, we aligned these separate embeddings through cross-modality contrastive learning. By contrasting samples across all possible pairs of modalities (neural activity, scenery frames, and local motion frames, see Supplementary Materials \ref{alltoall}), we generated a representation that encodes stimulus features shared between neural recordings and the movie. Cross-modality contrastive learning requires substantial computational resources, as effective training typically requires samples per class on the same order of magnitude as the total number of classes \citep{shahinfar2020many}. We drew hundreds of neural activity samples by randomly selecting subpopulations of neurons from the recorded population per sampling window, resulting in datasets of nearly half a million samples (see Supplementary Materials~\ref{method:pseudo} for detailed data preparation). In addition, cross-modality learning requires that three separately randomly initialized deep networks be trained simultaneously, compared to a single architecture in standard classification tasks. Due to these computational requirements, we demonstrate our method's capabilities using the first 400 frames of the movie. Because the autocorrelation of the scenery and the local motion features exhibits similar patterns throughout the movie (\ref{fig:decayscale}), our findings will generalize to the entire stimulus. 

The stimulus features identified in Results.~\ref{sec:movie} can occur anywhere in the visual field, but any individual V1 neuron samples only a restricted region of the stimulus. Consistent with this, V1 units recorded from a single animal collectively span only $\sim$30--40\% of the visual field \citep{Allen2019}. In addition, previous work \citep{Deitch2021} showed that the stimulus-relevant structure can be reliably recovered once their analysis includes neural populations of approximately 250--500 neurons. Therefore, in this study, we also created a ``pseudomouse'' by combining V1 units from multiple animals. Each pseudomouse includes data from a distinct set of five mice (see Supplementary Materials \ref{method:pseudo} for details), with aggregated receptive fields that span the full area of visual stimulus (see Supplementary Materials \ref{method:pseudo}). We then determined parameters for sampling a pseudomouse's neural activity based on known visual processing constraints. Because the discrimination of scenery features saturates only after $\sim$300~ms \citep{Resulaj2018}, we follow \citet{schneider2023cebra} and, for each frame $t$, extract samples of both neural activity and the corresponding movie segment from a window of 330~ms centered at $t$. To capture fast-changing local motion features, we implemented co-occurrence constraints at 33 ms, the frame rate of the movie. By enforcing these temporal co-occurrence constraints at the single-frame resolution across hundreds of frames simultaneously, our contrastive learning approach extracts an embedding of stimulus features and neural encoding fluctuating at 33 ms.

\begin{figure}
    \centering
    \includegraphics[scale=0.2]{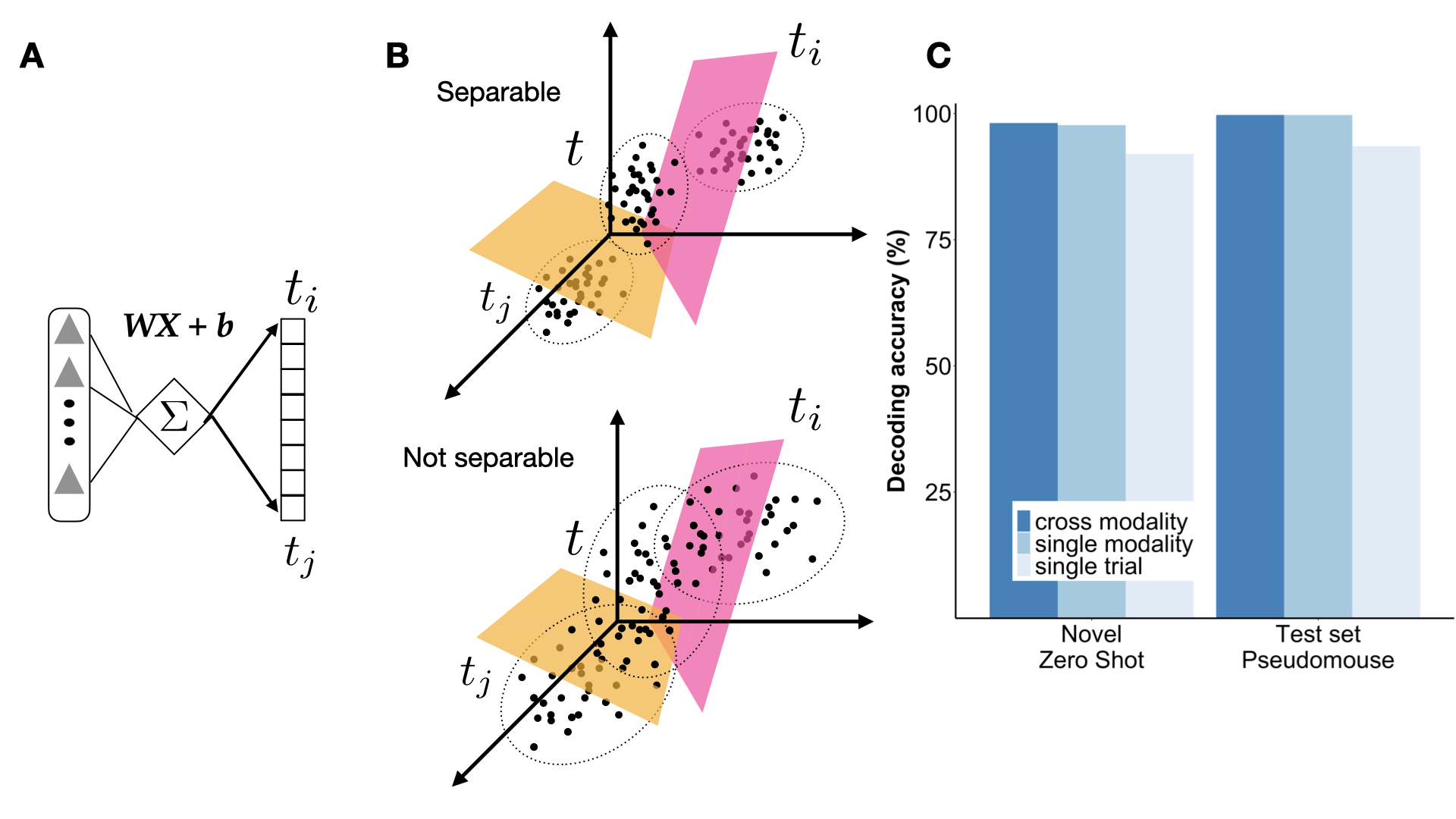}
\caption{\textbf{The learned embedding achieves near-optimal linear decoding performance after both single-modality and cross-modality learning phases.} A) We evaluate our embedding by training a linear decoder $WX+b$ to distinguish embeddings from the correct time frame $t_i$ versus all alternative time frames $t_j$. With $n$ total frames, each correct frame must be distinguished from $n-1$ alternative frames. The chance level for this task is $1/n$ with $n=400$, i.e., 0.25\%. B) The linear decoder learns multiple decision boundaries (hyperplanes) that partition the high-dimensional embedding space. In an optimal embedding (top schematic), neural activity from different time points forms distinct clusters, allowing these hyperplanes to create unique partitions for each time frame, enabling perfect linear separability. When embeddings overlap (bottom schematic), these hyperplanes cannot establish unique subspaces for each time point, resulting in decoding errors. C) Decoding performance comparison between the single-modality and cross-modality training phases, using either trial-averaged neural population responses (PSTHs, as held-out test sets) or single-trial activity. The embedding achieves approximately 99\% accuracy using PSTH data and 92-93\% using single-trial data when decoding frame numbers at 33 ms resolution.  Additional decoding results are available in Supplementary Materials \ref{supp:baseline}. Because we have 80,000 samples in our held-out test dataset, the standard errors of the decoding performance here are on the order of 0.1\% (not shown)}
\label{figure:timedecode}
\end{figure}

Since our contrastive learning is guided by temporal co-occurrence with single frame, a good sanity check is whether a linear decoder trained on the learned embedding can identify neural activity corresponding to a given frame and distinguish it from that of all other frames, where consecutive frames are 33~ms apart. As shown in Fig.\ \ref{figure:timedecode}, our models achieve nearly 99\% accuracy when decoding frame numbers (effectively time at 33 ms resolution) from held-out test data (Note that the chance level for this 1-vs-all decoding is 1/400, 0.25\%). Both single-modality and cross-modality trained models demonstrate this near optimal performance. As detailed in the Supplementary Materials \ref{supp:baseline}, even models that use only scenery features or only local motion features can decode time at this single frame resolution with at least 95\% accuracy. Our neural-based model (trained with only neural activity) achieves slightly better performance because it has access to both types of feature encoded in the V1 activity. The high decoding accuracy achieved after cross-modality training suggests that the learned embedding approximately minimizes the Bayesian decoding risk. Recent work on Bayesian risk minimization shows that learned embeddings that achieve minimal decoding risk exhibit support matching between domains \citep{Ruan2021, Dubois2021}. In practice, this makes our learned embedding a natural benchmark for quantifying representational drift, since changes in feature encoding across sessions necessarily decrease this decoding accuracy.

Beyond achieving near-optimal performance on the held-out test data, our learned embedding also generalizes well to other datasets. When applied to single trial neural activity, the embedding maintains approximately 93\% decoding accuracy within the animals included in the training, although it is trained only on the respective trial-averaged neural activity (PSTHs) (Fig.\ \ref{figure:timedecode}). It also achieves 92\% accuracy for a completely novel pseudomouse whose animals (neither PSTH nor single trial activity) were not included in training (see Supplementary Materials~\ref{method:pseudo}). This cross-subject generalization indicates that our model captures common stimulus features encoded across animals, as opposed to idiosyncratic features of the individuals' internal states. In addition, we benchmark our approach against conventional dimensionality reduction methods applied directly to neural population activity, including principal component analysis (PCA) and nonnegative matrix factorization (NMF) (see Supplementary Materials \ref{supp:baseline}). Our embedding substantially outperforms these methods, indicating that contrastive learning organizes stimulus-relevant features in a way that enables more efficient linear decoding than conventional unsupervised approaches. 

\subsection{Quantifying the magnitude of representational drift via changes in decoding performance}
\label{sec:diff}

\begin{figure}[!ht]
     \begin{subfigure}[t]{0.45\textwidth}
          \caption{}
         \includegraphics[width=0.85\textwidth]{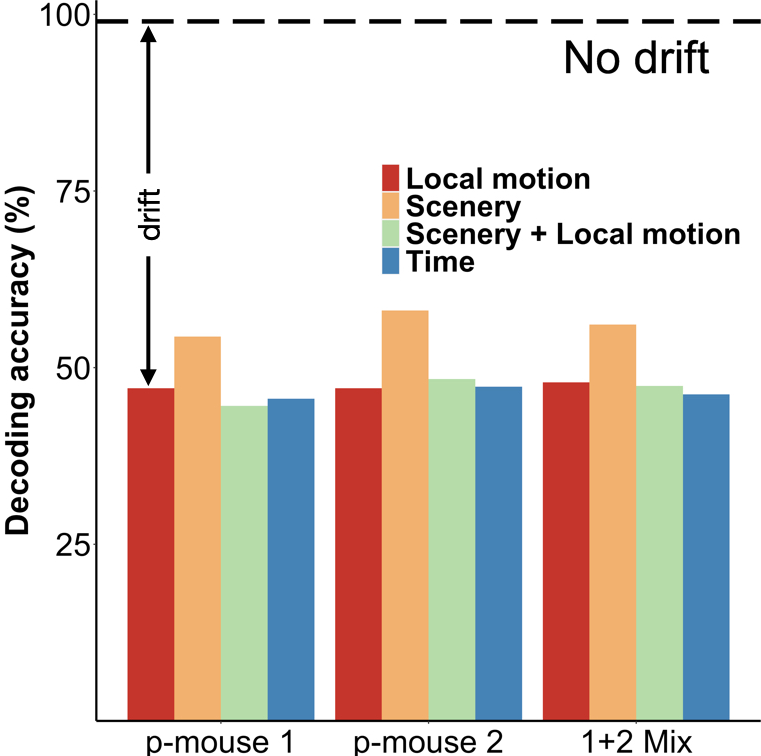}
         \label{fig:4A}
\end{subfigure}
     \begin{subfigure}[t]{0.45\textwidth}
        \caption{}
         \includegraphics[width=0.9\textwidth]{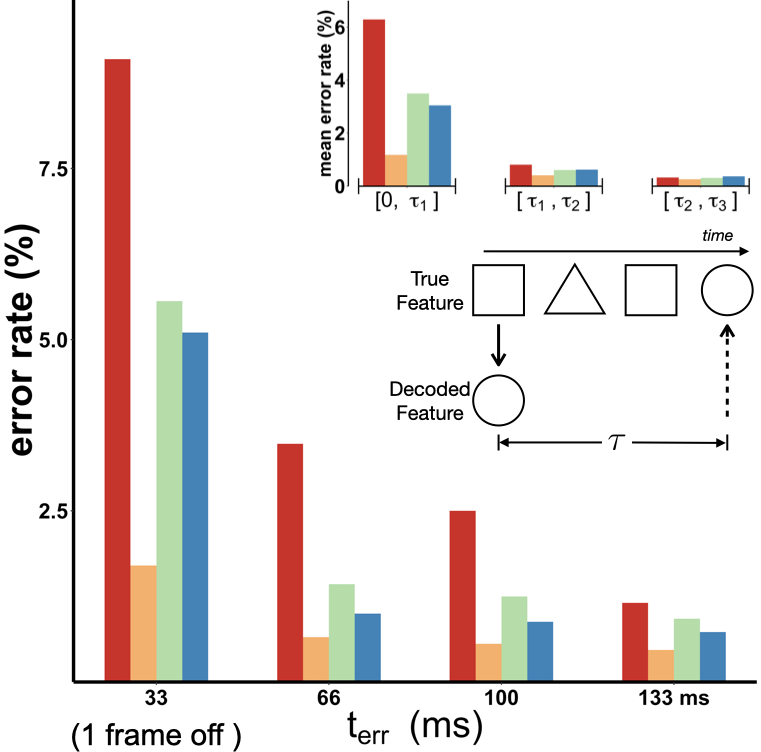}
         \label{fig:4B}
     \end{subfigure}
\caption{\textbf{Representational drift affects the encoding of stimulus features differently depending on their spatiotemporal statistics.} 
A) Overall effect of representational drift on stimulus feature decoding: We pretrained our model using neural activity from session 1 (training set) and froze the weights. We then trained linear decoders for four natural features using neural representations from session 1. Next, we generated two sets of neural representations: those without drift by mapping held-out test data from session 1 into the pretrained model, and those with drift by mapping neural activity from session 2 into the same model. The dashed lines show decoding performance without drift (approximately 99\% for all features, also shown in Fig.\ \ref{figure:timedecode}), while the bar plots show performance with drift, revealing significant degradation across all features. Note that standard errors of these accuracies scale inversely with sample size. Because we use 80,000 samples in both the held-out test set from session 1 (without drift) and the dataset from session 2 (with drift), the standard errors of the decoding performance are negligibly small (on the order of 0.1--0.3\%. Consequently, differences in decoding between stimulus features (e.g., Scenery vs Local Motion) are also significant ($p < 0.0001$).).
B) The drift rates between local motion and scenery features are different. Here we use the decoding error per frame as a proxy of this drift rate. This decoding error per frame is defined as the temporal distance between frames where the predicted and true features appear (see Supplementary Materials \ref{method:temporal}). The bottom inset illustrates how we compute this temporal distance. The decoded feature (round) is 3 frames (100 ms) away from the correct feature (square), thus, this decoding error has a temporal distance of 3. Here we plot the fraction of decoding errors with temporal distance $n$ as $n$ changes from  33 ms to 133 ms (1 to 4 frames off). The top inset shows the aggregated decoding errors within three time windows (by summing all frames within those windows): $(0, \tau_1)$ ($\tau_1 = 66$ ms), $(\tau_1, \tau_2)$ ($\tau_2 = 500$ ms), and $(\tau_2, \tau_3)$ ($\tau_3 = 1000$ ms, defined in Results. \ref{sec:movie}).}
\label{fig:decode1}
\end{figure}

Having established that our embedding can achieve near-optimal performance for decoding time in the movie at the single-frame resolution, we now explore its capacity to encode specific movie features. Our analysis in Results.~\ref{sec:movie} demonstrates that the single-frame resolution captures fluctuations from both local motion and scenery features operating at different timescales. Fig.\ \ref{fig:decode1} confirms that our model effectively differentiates these stimulus features we estimated in Results. \ref{sec:movie} and achieves near-optimal decoding accuracy (99\%) for all three types: local motion, scenery, and the joint features. This near-optimal performance establishes our embedding as a benchmark for quantifying subsequent representational drift. Since the model achieves such high feature decoding accuracy for the session it was trained on (results reported for holdout test set data), any changes resulting from representational drift must necessarily appear as decreases in decoding performance when applied to neural activity from a different recording session. This benchmark allows us to directly measure the magnitude of representational drift as the performance difference between sessions. Without a near-optimal baseline, quantifying representational drift would be challenging because it could lead to both decreases or increases in decoding accuracy.

Building on previous work \citep{Xia2021,Sadeh2022}, we trained linear decoders using neural activity from session 1 and evaluated their performance on both session 1 data (holdout test set) and session 2 data (from which we quantify representational drift). Training the model on the session 1 train set and applying it to session 1 holdout test set can be thought of as the case where there is no drift. We refer to these results as ``without representational drift''. In contrast, applying the model session 2 holdout test set is referred to throughout the text as ``with representational drift''. Our linear decoding design differs from previous work in two key aspects: First, we implemented a 1-vs.-all classification framework rather than the 1-vs.-1 (or so-called pairwise discrimination) used in \citep{Sadeh2022}. For each neural activity sample centered on frame $t_i$, our decoder determines whether it belongs to frame $t_i$ versus all other frames ($t_{1, 2, \cdots, i-1, i+1, \cdots, n}$, where $n$ is the total number of frames). The chance level here is 0.25\%, much lower than the 50\% in pairwise discrimination, where each frame is tested against only one other frame. Second, we decoded at single-frame temporal resolution (33 ms) instead of the 1-second windows. This finer resolution means that our decoder distinguishes between 30 times more possible labels. 

Although this fine-scale decoding task is more challenging, it enables us to track frame-to-frame changes in both local motion and scenery features, and thereby quantify how these features are affected by representational drift. This setup also makes contact with early work on information in spike trains that argues for using information about ``time in the trial'' as a proxy for stimulus features sampled by that repeated stimulus trace \citep{Brenner2000}. Within the 400-frame trial, the stimulus contains 42 scenery features, 69 local motion features, and 198 joint features. Stimulus features change on average every $\sim$2 frames. At this single-frame resolution, frame number serves as a proxy for the suite of complex features the movie at that particular time. Frames sharing similar scenery or motion content are expected to be confusable, regardless of where they are, while frames with distinct features should be distinguishable. The pattern of decoding errors thus reveals the temporal structure of stimulus feature dynamics, showing how quickly complex stimulus features decorrelate over time. Despite the increased complexity, our results remain consistent with earlier observations. This observed decoding degradation is more salient in our 1-vs-all setting than in the simpler pairwise discriminability. When we apply pairwise discrimination, we also observe persistent performance at single-frame resolution for session 2 data (see \ref{fig:driftdecode}), consistent with \citep{Sadeh2022}. We adopt 1-vs. all because all features in the movie stimulus fluctuate every 1 to 3 frames (see Results \ref{sec:movie}), making it a more appropriate test for tracking frame-to-frame changes in the features.


When applying our linear decoder trained on session 1 data to neural activity from session 2, we observed substantial performance degradation across all four features due to representational drift (see Fig.~\ref{fig:4A}. The overall decoding accuracy decreases by approximately 50\%, although this effect varied by features. The scenery features show the least degradation (retaining 56\% accuracy), while the local motion features exhibit the most (retaining only 47\% accuracy). To better understand these differences, we analyze how the decoding errors are distributed on different timescales. We quantify the error rate as a function of the temporal distance between the predicted and actual features. Specifically, for each temporal offset $\tau$ (measured in frames), we calculate the proportion of test samples whose decoding errors have a temporal distance of at most $\tau$. These are cases where the feature predicted by the linear decoder actually appears $\tau$ frames away from the frame corresponding to that test sample (shown in Fig.~\ref{fig:4B}). This analysis reveals that the decoding errors for local motion features are much more concentrated within $(0,\tau_1)$ than other features. Averaging the error rate within the interval $(0, \tau_1)$ reveals that local motion features have a mean error rate that is 5-6 times higher than those of the scenery features (inset) within $(0, \tau_1)$. The error rates for time and joint features fell between these extremes, reflecting their integration of motion and scenery information. Importantly, both the per-frame error rates and the aggregated decoding errors beyond $\tau_2$ converged to similar values across all features (the inset in Fig.\ \ref{fig:4B}). This convergence indicates that the approximately 9\% performance gap between local motion and scenery features comes primarily from errors within the shortest interval $(0, \tau_1)$, in which they show the largest disparity in error rates. Based on these observations, we hypothesize that drift changes the encoding of local motion features differently from scenery features. In the next section we explore how this difference arises from changes to the geometry of the embedding.

\subsection{Representational drift disrupts the decoding of local motion features through perturbations to the embedding geometry}
\label{sec:driftgeometry}

\begin{figure*}
\centering
     \begin{subfigure}[t]{0.30\textwidth}
     \caption{}
         \centering
         \includegraphics[width=0.9\textwidth]{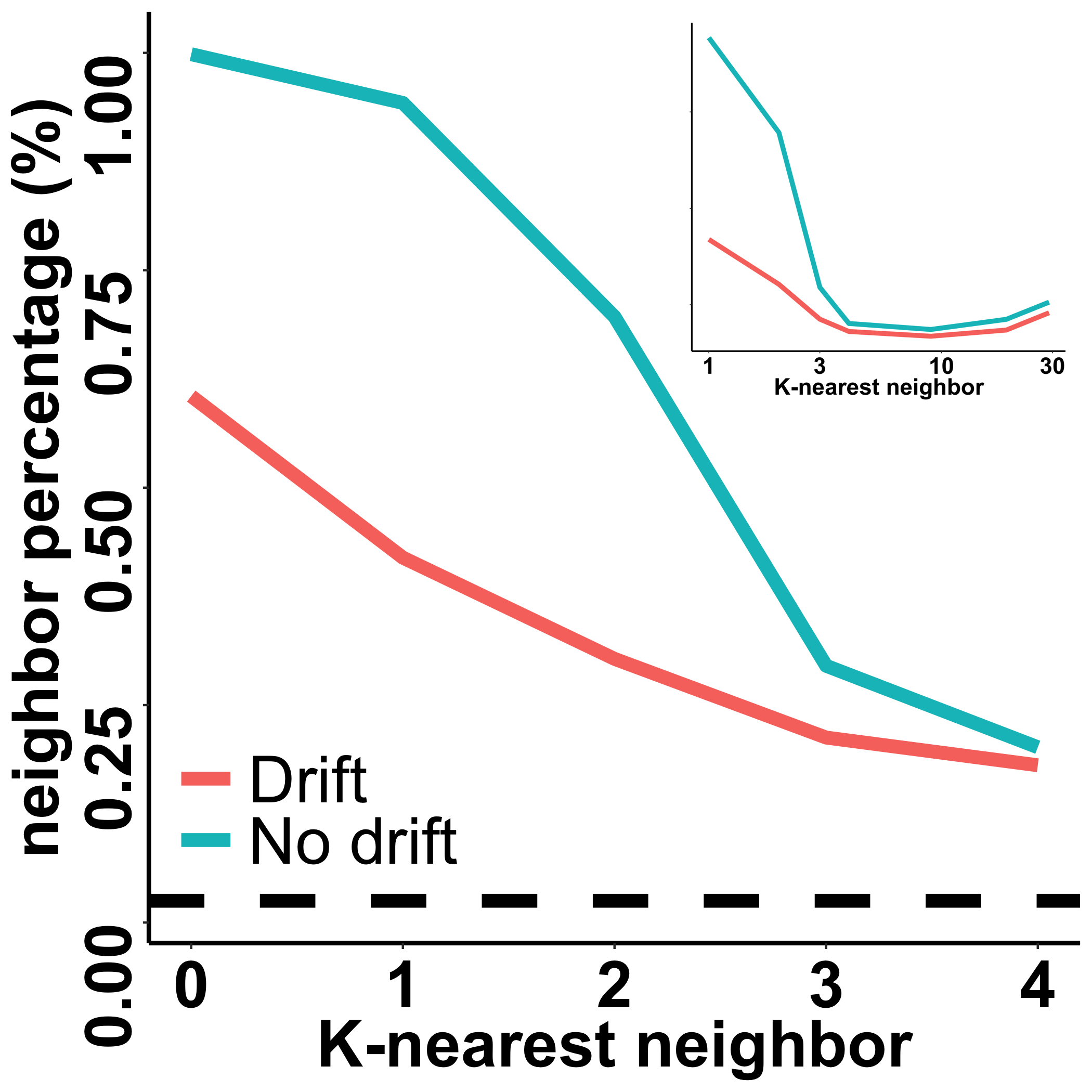}
         
         \label{fig:5A}
\end{subfigure}
\begin{subfigure}[t]{0.29\textwidth}
     \caption{}
         \centering
         \includegraphics[width=0.92\textwidth]{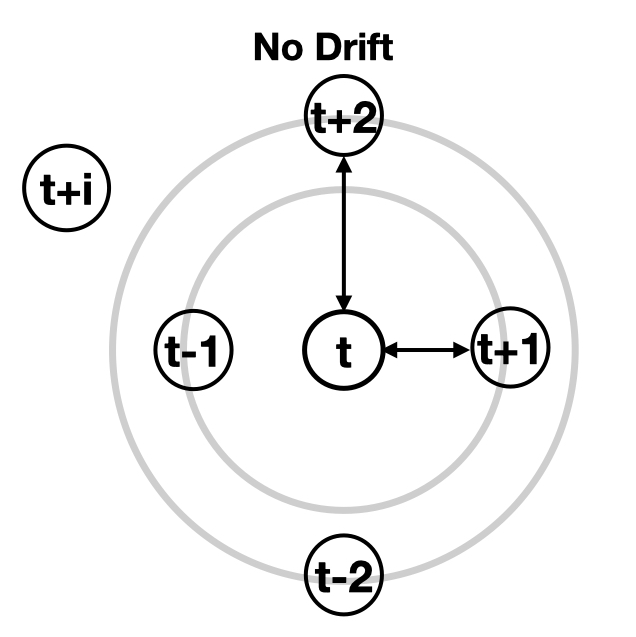}
         
         \label{fig:5B}
     \end{subfigure}
\begin{subfigure}[t]{0.29\textwidth}
     \caption{}
         \centering
         \includegraphics[width=0.92\textwidth]{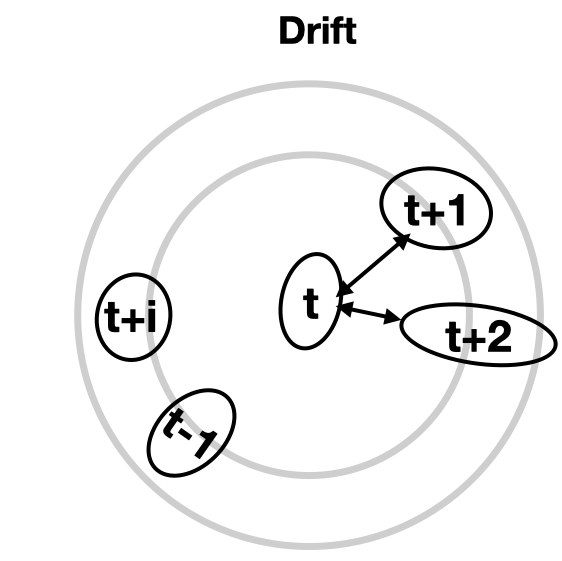}
         
         \label{fig:5C}
     \end{subfigure}

\caption{\textbf{Representational drift perturbs the local geometry needed for decoding local motion features.} 
\textbf{A)} The learned embedding without representational drift maintains a precise $K$-nearest neighbor structure for temporally adjacent frames within a local neighborhood ($k \in (0,4)$). The cyan line shows the proportion of test samples that maintain a local $k$-nearest neighborhood structure for $k \in (0,4)$. For a neural activity sample occurring at time $t$, its embedding belongs to the $k$-nearest neighborhood if its $k$-th nearest neighbor in the embedding space is a cluster corresponding to a time within $(t-k, t+k)$. At $k=0$, 99\% of test samples are correctly classified when using the class means as the linear decoder weights (see Supplementary Materials~\ref{appendix:NC}). At $k=1$, 96\% of test samples have their second nearest cluster (after their own time cluster) corresponding to either $t+1$ or $t-1$, preserving temporal adjacency. The magenta line shows how this same $k$-nearest neighborhood metric changes in the presence of representational drift, revealing substantial degradation. The dashed line indicates chance-level performance ($0.025$), and the gap between this line and the cyan/magenta curves reflects the preserved global geometric smoothness (see Supplementary Materials~\ref{sec:optimalgeometry}). The inset demonstrates that no significant difference exists when expanding the neighborhood beyond $k=4$, indicating that representational drift disrupts only local geometry. 
\textbf{B} and \textbf{C} Illustration of local geometry without drift \textbf{B} and with drift \textbf{C}. Without drift, features from different frames (e.g., $t$ and $t+i$) form well-separated clusters with regular shapes and consistent spacing. The $K$-nearest neighbor property ensures that temporally close neighbors (e.g., $t+1$ or $t+2$) are spatially closer than temporally distant ones (gray circles show the cluster of $t+2$ is approximately twice as far as the cluster of $t+1$). \textbf{C} shows that representational drift disrupts this organized structure by changing both cluster spreads and inter-cluster distances. }
\label{fig:NC4}
\end{figure*}

In this section, we investigate how representational drift changes the geometry of the learned embedding. We define geometry as the relative spatial arrangement of embeddings for neural activity in response to different movie segments. Our goal is to understand why representational drift disproportionately affects the decoding of local motion features compared to that of scenery features. Because local motion features change between adjacent frames (Results. \ref{sec:movie}), decoding them requires that the embeddings for neural activity occurring adjacent in time be separable. Our geometric analysis reveals that drift warps the embedding geometry such that embeddings for neural activity occurring close together in time overlap with each other, explaining the substantial decoding errors for local motion features (Fig.~\ref{fig:4A}).

To dig deeper into the features of the embedding that are warped by drift, we focused on two key geometric properties: global smoothness \citep{Caponnetto2007, Bordelon2022, Stringer2019a} and neural collapse \citep{Papyan_2020}. Global smoothness ensures that similar stimulus features are encoded near each other in the embedding space, creating a continuous stimulus manifold. In the large data regime, both theoretical investigations of kernel methods \citep{Caponnetto2007, Bordelon2022} and neural data analysis\citep{Stringer2019a} observed that such a smoothness is characterized by the asymptotic eigenspectrum decay of the learned embedding. Our analysis revealed that this global property remains intact despite representational drift (see Supplementary Materials \ref{sec:optimalgeometry}). This preservation of global structure suggests that the 5-6 times higher drift rate for local motion features cannot be explained by changes in global geometry. Given this finding, we investigate whether representational drift might disrupt local geometric properties. 

Neural collapse is a local geometric phenomenon that occurs towards the end of training in some high-performing networks. After achieving near-optimal decoding (see Results. \ref{sec:decode}), samples from the same class cluster tightly around their class mean, while different class means arrange in a geometrically regular separated structure \citep{Papyan2020}. In our context, the latent space embedding of neural activity at a specific time $t$ forms a distinct cluster separable from other embeddings of neural activity at other times (see Supplementary Materials \ref{appendix:NC}). Our embedding exhibits this neural collapse geometry in both train (\ref{suppfig:NC}) and test datasets (Fig.~\ref{fig:5A} and \ref{fig:newNC}). The combination of neural collapse and global smoothness creates a structured local neighborhood. Neural collapse ensures that the samples from each frame cluster around their distinct centroids, while global smoothness ensures that the samples from temporally adjacent frames ($t$ and $t+1$) are located near each other in the embedding space. Consequently, for a sample from time $t$, clusters corresponding to times $t\pm 1$ typically become its closest neighbors. This creates a precise ordering where clusters for temporally adjacent frames ($t\pm k$) automatically become the $k$-th nearest neighbors to time $t$ - a property that may be crucial for maintaining precise temporal relationships in the neural code.

Without representational drift (i.e.\ for session 1 holdout test set data), neural activity responding to nearby segments in the movie is mapped close together in the learned embedding. As shown by the cyan line in Fig.~\ref{fig:5A}, clusters corresponding to frames $t-1$ or $t+1$ serve as the first nearest-neighbor for 96\% of the neural activity samples over time $t$. For 75\% of the samples, their second nearest-neighbors correspond to frames $t\pm 2$. This structured relationship rapidly diminishes beyond 2-3 frames, creating a temporal precision in the representation space that aligns with the 33-100 ms autocorrelation window of local motion features. Representational drift substantially weakens this precise local structure (magenta line in Fig.\ \ref{fig:5A}). For session 2 data, only 61.5\% of the neural activity we sampled at time $t$ is closest to the centroid of time $t$ (instead of the centroids of other times) in the embedding space. This is a reduction of 38.5\% resulting from the representational drift. More critically, only 41.9\% and 30.3\% of the test samples maintain $t\pm 1$ and $t\pm 2$ as their first or second nearest neighbors, respectively. This geometric perturbation disrupts the linear separability of the embeddings of neural activity for adjacent frames, particularly within the critical 4-frame window (approximately 33-100 ms). 

In particular, the inset of Fig.~\ref{fig:5A} reveals that when expanding $K$ to 10-30 frames (up to 1 sec), the proportions of samples following the neighborhood structure become nearly identical with and without representational drift. This observation aligns with our findings in Fig.\ \ref{fig:decode1}, where drift rates between local motion and scenery features differ dramatically within 2-3 frames, but converge beyond 10 frames. Although both local motion and scenery features experience substantial decoding degradation due to representational drift, the preservation of geometric structure beyond 10 frames helps explain why this degradation is less severe for slow-varying scenery features (with 500-1000 ms autocorrelation) compared to rapidly changing local motion features.

Figs.~\ref{fig:5B}--\ref{fig:5C} (see also \ref{fig:newNC}) illustrate how drift disrupts the local geometry required to decode fast-varying local motion features. Without drift, the embeddings of time $t$ and its immediate neighbors ($t\pm1$, $t\pm2$) form well-separated and regularly arranged clusters; with drift, these nearby embeddings deform and partially overlap, increasing errors when linearly discriminating frame at $t$ from frames at adjacent time steps. Consistent with this, drift perturbs the neural collapse geometry by increasing the variability in centroid angles and weakening the alignment between classifier weights and class means (Supplementary Materials~\ref{appendix:NC}). These distortions are concentrated within a $\sim$4-frame neighborhood (33--100 ms), matching the autocorrelation timescale of local motion features. Consequently, even retraining a linear decoder fails to correctly differentiate local motion features occurring in close temporal proximity (see \ref{fig:driftdecode}), confirming that drift imposes a limit on the linear separability of features at this fine timescale.

\section*{Discussion}

We observe representational drift in the encoding of stimulus features that fluctuate over a temporal range spanning approximately one order of magnitude (33--500~ms). Alongside previous findings that odor-evoked responses in the primary olfactory cortex drift continuously \citep{Schoonover2021}, our results suggest that drift in sensory systems generally changes the encoding of external stimuli. This change undermines the reliability of any fixed linear mapping between neural activity and stimulus variables. Combined with the observation that drift in V1 over multiple days tends to stabilize rather than gradually change the encoding map \citep{Aitken2022}, our findings further support the idea that representational drift in the sensory cortex differs from that observed in the hippocampus, sensorimotor, or motor circuits, where drift often appears as a cumulative change in the population code \citep{Driscoll2017, Rule2020}. Together, these findings raise the possibility that representational drift may serve different functional roles across brain systems and timescales.

We also demonstrate that representational drift occurs heterogeneously across sensory features, depending on their intrinsic timescales. By comparing a fast-varying local motion feature with a slower-varying scenery feature, we find that drift changes the encoding of the local motion features at a rate 5--6 times higher than that of scenery features. This suggests that representational drift is not a uniform process acting on the population code; instead, the stability of feature encoding may depend on how rapidly those features fluctuate. Combined with previous observations that slower timescales (on the order of 1~second) are preferentially associated with encoding internal states \citep{Sadeh2022}, our findings raise the possibility that feature dynamics serves as an organizing principle of the V1 population code. Consistent with the proposals that V1 supports a multi-dimensional factorized representation of sensory and internal variables \citep{Buzs_ki_2019}, features with distinct temporal dynamics can be assigned to independent representational dimensions. Such separation may insulate the encoding of one feature from drift in another, preserving the robustness of the neural code. Notably, the two features we examine map onto distinct ethological functions: local motion features (e.g., looming or localized movement) are critical for triggering fast escape or predator-avoidance responses, whereas slow-varying scenery features support spatial orientation and foraging (e.g., recognizing locations or food sources). When representational drift affects the encoding of motion detection, it minimally impacts the encoding of scenery textures. Future experiments that manipulate stimulus dynamics across timescales could test whether this modular organization constitutes a general design principle for reliable yet flexible visual processing in complex natural environments.

In addition, we suggest that the representational drift in V1 may change the nonlinear encoding of simulus features, rather than simply rotating the encoding within a fixed linear subspace. This interpretation is based on two observations: first, retraining linear decoders after drift recovers only partial decoding performance (see Table \ref{tab:perf}); and second, decoding based on principal components performs substantially worse than decoding from our learned (nonlinear) embedding (see Supplementary Result \ref{supp:baseline}). These results suggest that drift changes not only the orientation of the neural code, but also its geometry in ways that linear decoders cannot fully recapture (see Results \ref{sec:driftgeometry}). This contrasts with previous experimental and theoretical work showing that drift can be effectively modeled as a rotation of the encoding map. For example, in the piriform cortex, a decoder trained on day~1 dropped to chance on day~32, but a new decoder trained on later neural activity still accurately classified odor identity \citep{Schoonover2021}. This successful retraining implies that stimulus information remained in a linear subspace, with drift acting as a rotation between the old and new decoder weights. In other words, drift preserved the linear subspace encoding the stimulus, changing only the orientation of that space. Similarly, in spatial navigation tasks, modest adjustments to the linear decoder weights maintained stable decoding for many days, implying that the variables relevant to the task remain accessible via linear transformation despite ongoing drift \citep{Driscoll2017, Rule2020, Rule2022} (e.g., a diffusion process \citep{Qin2023, Natrajan2025, Morales2025, Pashakanloo2023}). By contrast, in our case, retraining a linear decoder after drift yielded only partial recovery, suggesting that the embedding of neural activity had moved outside the original linear subspace. Because our decoding framework is designed to track specific stimulus features present in the movie, we hypothesize that this discrepancy arises from the complexity of natural scene features. Naturalistic stimuli are likely to drive higher-dimensional, less stereotyped patterns of activity in V1, and their encoding may involve nonlinear transformations that are more susceptible to representational drift. Moreover, previous studies often decoded abstract variables such as temporal context or spatial position—quantities that may be encoded stably over time and are less likely to require nonlinear transformations. In contrast, features in natural scenes are dynamic and high-dimensional, placing greater demands on the flexibility of sensory representations and potentially amplifying nonlinear changes under drift.

What mechanisms could plausibly drive the representational drift we observe in V1 over a short interval of 90 minutes between recording sessions? This timescale is much shorter than the day-to-week window typically associated with intrinsic excitability \citep{Delamare2024, Climer2025, Haimerl2025}, structural rewiring processes such as dendritic spine turnover \citep{Attardo2015} or synaptic scaling \citep{Turrigiano1998}. Any explanation must therefore invoke forms of plasticity that operate on the scale of minutes to hours. A recent computational study by \citet{Morales2025} offers a concrete candidate mechanism. Using a model of representational drift in piriform cortex, they considered synaptic changes arising from two distinct processes: slow, spontaneous synaptic fluctuations and rapid spike-timing-dependent plasticity (STDP) engaged during repeated stimulus exposure. While STDP stabilizes representations over long timescales, the model revealed that it can instead contribute to drift during an early exposure phase lasting minutes to hours. In this regime, each stimulus presentation drives STDP-mediated synaptic updates that push network activity toward a stimulus-specific attractor. When the network is initially misaligned with this attractor, these updates can be large, producing substantial changes in encoding over short timescales. Notably, this early exposure window is compatible with the 90-minute separation between our recording sessions. 

STDP is also a plausible mechanism for the nonlinear changes in feature encoding that we observe in V1. Timing-dependent LTP/LTD rules have been characterized at neocortical excitatory synapses in visual cortex, establishing that spike timing can reshape synaptic efficacy on behaviorally relevant timescales. Mechanistically, these timing-dependent changes are gated by identifiable coincidence detectors and retrograde signals, including NMDAR-dependent components for potentiation and endocannabinoid/CB1R-linked pathways for depression \citep{Sjostrom2004, De_Pasquale2014, Feldman2012}. Such biochemical gating makes STDP both rapid and state-dependent, allowing repeated movie stimulus to induce substantial synaptic updates over minutes to hours. This picture is also consistent with our observation that drift cannot be explained by a simple linear remapping. 

Rapid plasticity in inhibitory circuitry may also result in nonlinear distortions of representational geometry. Visually induced endocannabinoid-dependent long-term depression at inhibitory synapses has been proposed to reshape GABAergic transmission in developing visual cortex, and interneuron subclasses are known to regulate when and where cortical plasticity is expressed \citep{Jiang2010, Van_Versendaal2016}. Changes in inhibition that selectively suppress overactive responses can, in principle, compress or warp the representational space rather than simply translate it, aligning again with our observation that drift is not well captured by a purely linear remapping. 

Finally, the magnitude of drift we observe, approximately a 50\% degradation in decoding accuracy, is compatible with experimental studies showing that even fast homeostatic mechanisms acting on inhibitory circuitry need many hours to regulate network dynamics \citep{Ma2019, Hengen2013, Wen2024}. Some previous theoretical work focused on how a single mechanism may contribute to drift \citep{Pashakanloo2023, Haimerl2025}. When they included multiple mechanisms \citep{Rule2022, Qin2023, Natrajan2025}, they usually kept those mechanisms operating at a single timescales. Our findings extend this line of thinking by suggesting that the timing-dependent interactions between synaptic plasticity and homeostatic regulation may differentially reshape the encoding of complex naturalistic features. This is a question interesting for further investigation.

An important open question is whether the representational drift we observe in V1 impacts how downstream circuits decode stimulus features from V1. If we assume that the animal’s ability to respond to stimulus features remains stable despite drift, what compensation mechanisms are available in downstream circuits? For slow-varying scenery features, a straightforward strategy is to leverage population averaging or redundancy, e.g., pooling from a large V1 population \citep{Stringer2019}. By doing so, a downstream circuit could potentially maintain a stable readout by pooling over many V1 neurons and re-weighting them over time. This pooling recalibrates and averages out small changes from drift. Similarly to a spatial navigation task \citep{Rule2020}, scenery features could be used for foraging, and it is reasonable to hypothesize that a similar mechanism that periodically updates its weights to counteract drift also applies here.

For local motion features, the search for compensation mechanisms is more challenging and uncertain. After a 90-minute interval, drift causes nearly half of local motion features to be misclassified as features occurring 33 to 100 ms away, just one to three frames apart. This timescale of error poses a fundamental challenge for known compensation mechanisms. The visual system typically requires integration times of approximately 80 ms or more to reliably discriminate complex patterns or textures \citep{Resulaj2018}. This temporal mismatch means that mechanisms like pooling, which can work effectively for slow-varying features, might be ineffective or even counterproductive for fast-varying features. Pooling across more neurons can stabilize the encoding of slow-varying features by averaging out the noise, but this approach may introduce additional delays that would further degrade the encoding accuracy of fast-varying features. Our geometric analysis provides additional insights on why compensation is challenging. Representational drift nonlinearly warps the local geometry such that it limits linear separability of fast-varying visual features. Error tolerance coding seems like the most straightforward choice:  behavioral stability may not require the full decoding accuracy available without drift. For example, V1 can discriminate orientation changes of $\sim 0.1^\circ$ in gratings, while behaviors only need discrimination of $\sim 5^\circ$ \citep{Stringer2019}, providing a 50-fold error margin. This is unlikely to be the only mechanism given the extreme temporal precision of motion-guided arrest/escape behaviors driven by the superior colliculus (SC) that operate on timescales as brief as 1-2 ms \citep{Liang2015}.

The pathways in SC could perform a comparison and integration of stable vs.\ drifting inputs: SC receives direct input from the retina as well as input from the visual cortex, converging onto the same neurons. The retinal input provides a largely stable, hard-wired representation of certain stimulus features (for example, a looming shadow will always send inputs to SC circuits directly from the retina) \citep{Gale2016,Kuhn2025}. The V1 input, on the other hand, can modulate SC neurons, which effectively acts as a dynamic gain or context signal. If drift changes the input from V1, the SC could recalibrate, relying momentarily more on direct retinal input. During minutes or hours between recording sessions, the SC could also adjust synaptic weights through fast plasticity mechanisms (STDP is also a candidate mechanism here), effectively learning the new correspondence between V1 activity and the ground truth provided through the retina. This provides interesting future directions to explore whether simultaneous recording from V1 and SC (or reading out behavior) is possible \citep{Savier2019, Zhao2014}. Do SC-driven behaviors remain stable while V1 representations drift? Given V1's modulatory role? How do SC neurons recalibrate their readout of drifting cortical inputs and combine them with stable retinal inputs? 

Our mathematically principled, weakly supervised, cross-modality contrastive learning method opens new avenues for investigating neural coding under complex, naturalistic conditions. As neuroscience increasingly embraces the richness of natural behaviors and stimuli, previous methods face a fundamental challenge: ground-truth ``features'' driving neural activity in naturalistic settings are often elusive to define \textit{a priori}. Our approach circumvents the need for predefined features by leveraging co-occurrence, based on the intuition that neural activity responding to the same external stimulus should encode a common set of features. While the learned embedding reflects the shared features between the stimulus and neural responses, its expressiveness ultimately depends on the richness of the stimulus and the resolution of the recorded activity. Given the flexibility of modern machine learning, our method is well suited to capture the high-dimensional temporally structured relationships inherent in naturalistic experiments. 

\setcounter{figure}{0}
\setcounter{section}{0}
\setcounter{subsection}{0}
\setcounter{table}{0}
\renewcommand{\figurename}{}
\renewcommand{\thefigure}{Supp Fig.~\arabic{figure}}

\section*{Supplementary Materials}

\subsection{Obtaining scenery/local motion features of the movie based on hierarchical clustering}
\label{supp:clustering}

We independently created clustering hierarchies for the 400 scenery frames and their corresponding local motion frames using agglomerative hierarchical clustering \citep{scikit-learn}. Since each frame contains $304 \times 608 = 184{,}832$ pixels, we first reduced dimensionality by passing all images through a ResNet50 pretrained on ImageNet, using the resulting 2048-dimensional feature vector as a proxy representation for each frame. We used ResNet50 pretrained on ImageNet because its training on diverse natural images produces features sensitive to both spatial structure (edges, textures, shapes) and color. Grayscale scenery frames activate the structural features, while local motion frames—which encode motion direction as color (e.g., leftward motion in one hue, rightward in another)—activate the color-sensitive features that distinguish between these directional categories. We then computed pairwise Euclidean distances between all frames in this feature space to form a distance matrix. The clustering algorithm takes a bottom-up approach: it begins with each frame as an individual cluster and iteratively merges pairs of clusters, with each merge greedily minimizing total within-cluster variance using Ward's criterion \citep{Ward63}. The algorithm performs 399 merging steps, terminating when all frames belong to a single cluster.

\begin{figure}
     \begin{subfigure}[t]{0.45\textwidth}
     \caption{}
         \centering
         \includegraphics[width=0.9\textwidth]{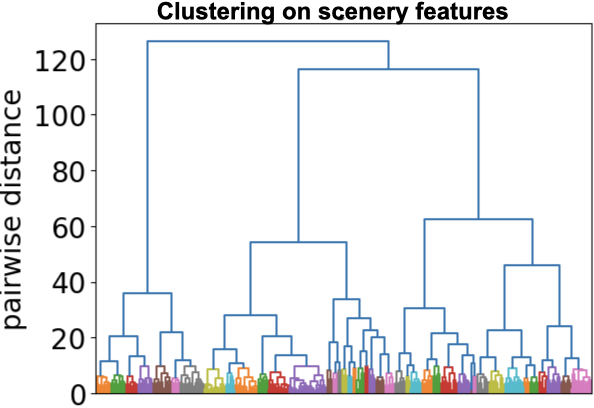}       
         \label{fig:scenecluster}
\end{subfigure}
     \begin{subfigure}[t]{0.45\textwidth}
     \caption{}
         \centering
         \includegraphics[width=0.9\textwidth]{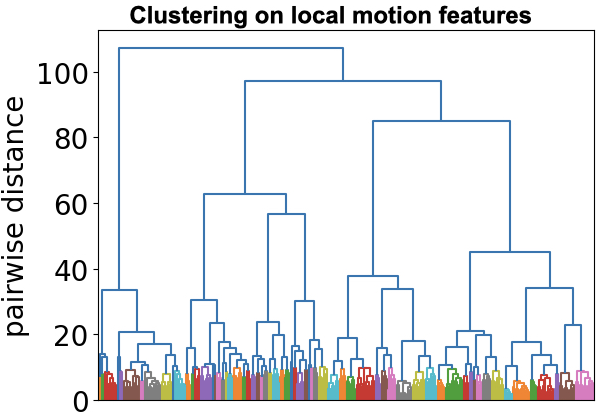}
         \label{fig:flowcluster}
     \end{subfigure}
\\
     \begin{subfigure}[t]{0.45\textwidth}
     \caption{}
         \centering
         \includegraphics[width=0.9\textwidth]{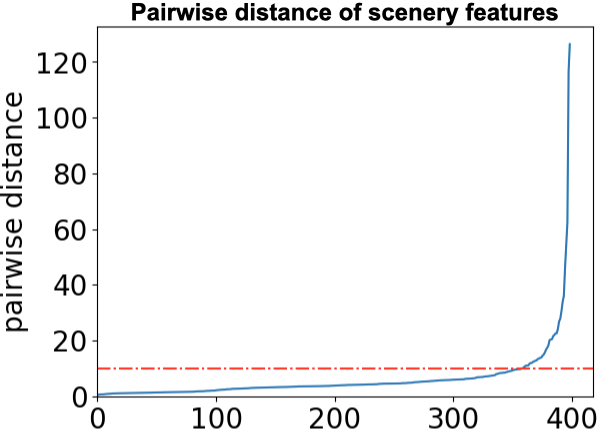}       
         \label{fig:scenedist}
\end{subfigure}
     \begin{subfigure}[t]{0.45\textwidth}
     \caption{}
         \centering
         \includegraphics[width=0.9\textwidth]{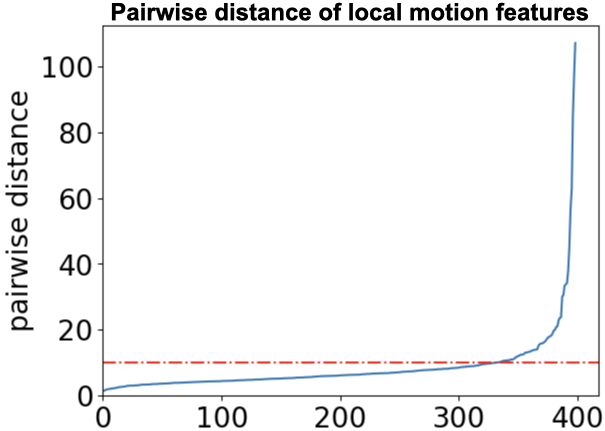}
         \label{fig:flowdist}
     \end{subfigure}
\caption{Hierarchical clustering on scenery frames and local motion frames, based on the greedy agglomerative clustering algorithm available from sklearn \citep{scikit-learn}. We use the \enquote{ward} method that minimizes the variance of the clusters being merged. A) The resulting clustering (shown as colored clusters) for scenery frames. B) Same as A), but for the local motion frames. C) The distance (y axis) as we merge frames (shown as x axis). We show the threshold as the red dashed line. We choose this threshold based on the knee of the respective distance function (between nodes in the corresponding children nodes of the agglomerative clustering). D)The same as C), but for the local motion frames. In addition, the joint feature combining clusters in C) and D) retains 84\% of the entropy compared to time. This suggests that these clusterings capture substantial decodable variability in the natural movie stimulus.}
\label{fig:allcluster}
\end{figure}

At each merge, the algorithm outputs a distance between the two clusters being combined. This distance quantifies cluster dissimilarity and increases monotonically as the algorithm progresses up the hierarchy. We generate discrete feature distributions by thresholding the hierarchy based on this distance: all clusters immediately below the threshold are treated as distinct features. In this work, the local motion frame at time $t$ is computed as the difference between scenery frames at $t$ and $t+1$. Lower thresholds yield many clusters with few frames each (high-entropy distributions), while higher thresholds group many frames together (low-entropy distributions). We observed that clustering distances grow slowly at first and increase exponentially toward the end, and selected thresholds near the knee of this curve (Fig.~\ref{fig:scenedist} and Fig.~\ref{fig:flowdist}).

\subsection{Settings of the weakly supervised contrastive learning}
\label{alltoall}
\begin{figure}[!ht]
     \begin{subfigure}[t]{0.45\textwidth}
    \caption{}
         \centering
         \includegraphics[width=0.99\textwidth, height = 0.8\textwidth]{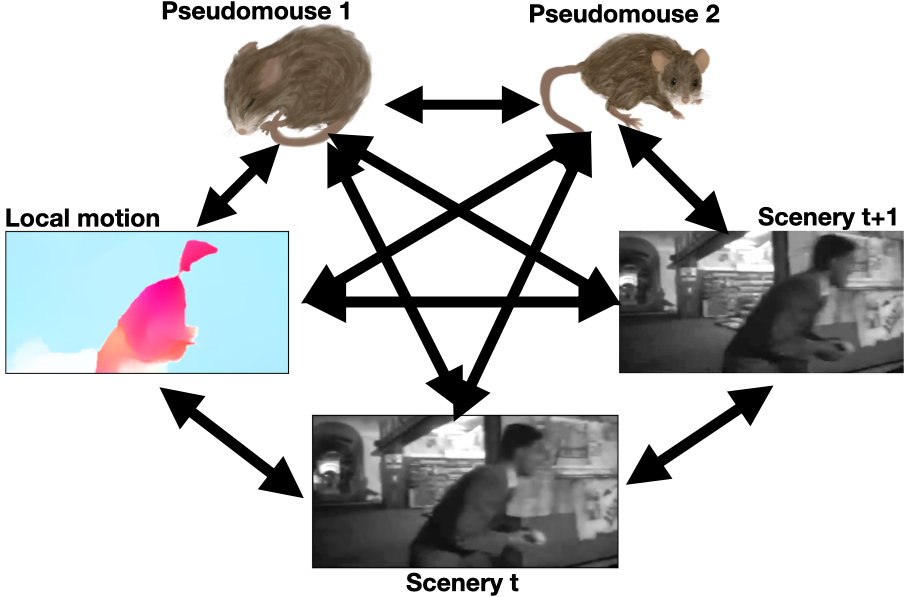}       
         \label{fig:allpairs}
\end{subfigure}
     \begin{subfigure}[t]{0.45\textwidth}
     \caption{}
         \centering
         \includegraphics[width=0.8\textwidth, height = 0.8\textwidth]{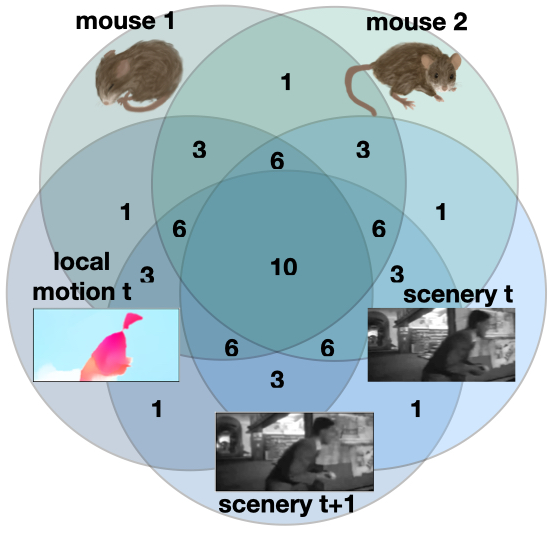}
         \label{fig:venn}
     \end{subfigure}
\caption{A) All-to-all contrastive learning during the cross-modality phase. There are five views in total: two from pseudomice 1 and 2, two from frames t and t+1, and one for the local motion frames at $t$. Single-modality phase: neural backbone is trained by contrasting pseudomice 1 vs.\ 2; scenery and local motion backbones are trained by contrasting their respective views. B) Venn diagram illustrating the all-to-all contrastive learning structure from A). Because all possible pairwise contrasts are being used in the cross-modality phase to refine all three backbones simultaneously, the information partition that is shared by all pairs gets the most emphasis. This includes the common sources of variation between neural responses and the external movie stimulus. }
\label{fig:allpairs}
\end{figure}

This section describes how we performed contrastive learning in the cross-modality phase and explains why this approach emphasizes information shared across all modalities. \ref{fig:allpairs} shows the 10 possible pairs among five views: two pseudomice, two scenery frames (at $t$ and $t+1$), and one local motion frame. We weight all contrastive pairs equally when computing the loss. As illustrated in \ref{fig:venn}, the number of arrows indicates relative emphasis: the partition shared by all modalities receives 10 arrows (highest weight), while the partition shared by the two pseudomice and local motion features receives 3 arrows only (mouse 1 with local motion, mouse 2 with local motion, and mouse 1 with mouse 2). 

\begin{figure}
     \begin{subfigure}[t]{0.45\textwidth}
    \caption{}
         \centering
         \includegraphics[width=0.8\textwidth, height = 0.8\textwidth]{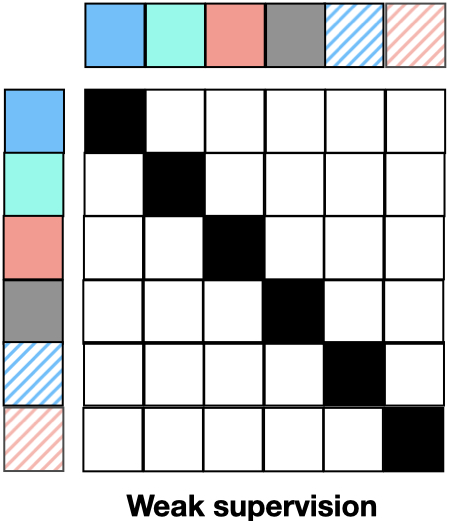}       
         \label{fig:weak}
\end{subfigure}
     \begin{subfigure}[t]{0.45\textwidth}
     \caption{}
         \centering
         \includegraphics[width=0.8\textwidth, height = 0.8\textwidth]{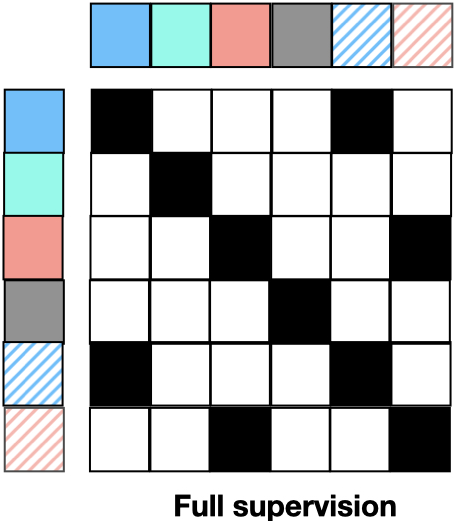}
         \label{fig:full}
     \end{subfigure}
\caption{Weak supervision (co-occurrence supervision) vs.\ full supervision (so-called supervised learning in a conventional setting). A) The weak supervision constructs positive pairs from distinct views (illustrated here as column and row) only if they are sampled into the same tuple (in our case, this is determined by whether they co-occur within the same sampling window). Otherwise, they are regarded as negative pairs despite sharing the same label (for example, the blue block and the blue stripe share the same label, but they are different tuples in the minibatch, so weak supervision treats them as negative pairs). B) In the case of full supervision, labels are included in contrastive loss \citep{Khosla2020}. All pairs of views that share the same label are considered positive. As a result, the blue block and blue stripe form a positive pair (shown as an off-diagonal black block).}
\label{fig:supervision}
\end{figure}

This section illustrates the key difference between weak supervision guided by temporal co-occurrence (our method, motivated by CLIP \citep{CLIP2021}) and full supervision (conventional supervised learning). In our weak supervision method, we do not use label information to define positive pairs. Instead, samples become positive pairs solely based on temporal co-occurrence: neural activity and visual stimuli sampled within the same temporal window form positive pairs, regardless of their labels (\ref{fig:weak}). In full supervision, explicit labels directly determine positive pairs: all samples sharing identical labels are treated as positive pairs regardless of their temporal relationship, producing off-diagonal positive pairs as shown in \ref{fig:full}.

\subsection{Dataset and Training details}
\label{method:pseudo}
 
We focus on the 24 recording sessions included in the Allen Visual Coding dataset collected with Neuropixels probes, each containing 30 trials. This larger number of trials (compared to 10 trials in other sessions) provides more accurately sampled mean activity traces (PSTH). For each time $t$ (frame number $t$), we sampled the respective mean neural activity using a 330-ms window centered on $t$. Since neural activity is sampled at 30 Hz, each V1 unit is represented by a 10-dimensional vector. We then aggregated five to six animals to produce a pseudomouse (see Table~\ref{tab:pseudomouse} for details). To generate a single sample of neural activity from a pseudomouse, we randomly sampled 300 units of the neural population of each pseudomouse, which contains 340–390 V1 units.

We prepare the dataset for contrastive learning following well-known observations for multi-class classification. Given that we have hundreds of frames whose frame numbers or features serve as class labels, we ensured adequate sampling by generating hundreds of neural activity patterns per frame. Limited by available multi-GPU computational hours, our analysis focuses on the first 400 of the 900 available frames in the movie. This subset captures sufficient diversity in visual features available in the movie (demonstrated in Supplementary Materials \ref{autocorr2}) while maintaining the necessary sample-to-class ratio for reliable deep neural network training \citep{cho2015much, shahinfar2020many}.

For each frame, we randomly sampled 600 neural activity patterns within an individual pseudomouse, using 500 for training/validation and reserving 100 for testing. Across 400 frames and two pseudomice, this yields 400,000 training samples and 80,000 test samples in total. Contrastive learning uses the full 400,000 training samples from both pseudomice. For linear decoders, if we refer to a linear decoder or its performance as ``p-mouse 1'' or ``p-mouse 2'', it uses data from that pseudomouse only (200,000 train / 40,000 test). Otherwise (e.g., ``1+2 mix'' in Fig.~\ref{fig:decode1}), the decoder uses pooled data from both pseudomice (400,000 train / 80,000 test).

We convert population neural activity into 2D tensors, following an approach originally developed for financial time series \citep{Wang2015, Barra2020} and later adapted to biomedical applications \citep{Fan2023, Yoon2025}. These works introduced the Gramian angular field, which represents a 1D time series as a 2D image by encoding the pairwise temporal correlation between values at times $t_i$ and $t_j$, so that 2D convolutional networks can capture temporal dependencies within time series. In our previous work \citep{wang2022learning}, we modified this construction by placing the raw data along the diagonal and the Pearson correlation between activity at $t_i$ and $t_j$ off the diagonal, and showed that this conversion efficiently reconstructs natural scenes from population retinal activity. We adopt the same approach here: each unit's 10-dimensional PSTH is converted into a $10 \times 10$ matrix whose diagonal contains the PSTH values and whose off-diagonal elements are the correlation of mean firing rates between times $t_i$ and $t_j$. A randomly sampled 300-unit activity tensor therefore has $300 \times 100 = 30{,}000$ entries, which we reshape into 30 channels of 1{,}000 entries each.

\begin{table}[]
    \centering
    \begin{tabular}{|c|c|}\hline
       Pseudomouse 1 &766640955, 767871931,768515987,771160300,771990200, 739448407\\\hline
       Pseudomouse 2 &774875821,778240327, 778998620,779839471,781842082,786091066\\\hline
       Novel Pseudomouse &787025148,793224716,794812542,816200189\\
       &821695405,829720705,831882777\\\hline
    \end{tabular}
    \caption{Session IDs we use to create pseudomice. Within these sessions, we choose those units of primary visual cortex in our analysis. Each session is an individual animal. There are 24 sessions in total with 30 trials for both session 1 and session 2. We group these sessions into three pseudomice randomly. The resulting pseudomice contains 340-390 V1 units each, similar to the set up used in \citep{schneider2023cebra}.}
    \label{tab:pseudomouse}
\end{table}


We extended the publicly available code at \url{https://github.com/HobbitLong/SupContrast} for multiview contrastive learning. All hyperparameters were set to default values. We used the \enquote{SimCLR} option throughout training, which implements temporal co-occurrence as the basis for positive pairs. All models used ResNet50 as the backbone architecture. Training proceeded in two phases: single-modality training for 300 epochs, followed by cross-modality training for an additional 300 epochs. Cross-modality training simultaneously optimizes three ResNet50 backbones (neural activity, static scenery, and local motion) on our 400,000-sample dataset. We used an 8$\times$A100 GPU cluster with a batch size of 2048. After training, we froze the model weights and trained linear classifiers on the embedding of frozen neural backbone. We then report the top-1 accuracy as the decoding performance.

\subsection{Calculation of temporal distance between the predicted and true feature}
\label{method:temporal}
To characterize the temporal distribution of feature decoding errors, we compute what we call a frame-specific error rate. For all samples in the test set that occur in frame $t$, we identify the proportion in which the linear decoder makes an incorrect feature prediction. If the nearest occurrence of the predicted feature is at frame $t+n$, the temporal decoding error is $|n|$ frames. We calculate the proportion of incorrectly decoded samples that have a temporal error of exactly $n$ frames, and plot this proportion as a function of $n$ in Fig.\ \ref{fig:4B}. This reveals the temporal intervals into which feature decoding errors predominantly fall. For example, whether the predicted features typically correspond to the actual features that occur less than 100 ms away (within the autocorrelation timescale $\tau_1$ of the local motion features). To compute the error rate within a temporal range of $\tau$ frames (Fig. 4B inset), we average the error rates for all values of $n$ within that range.

\subsection{Theoretical framework for interpreting near-optimal decoding and characterizing embedding geometry}
\noindent \subsubsection*{Interpretation of near-optimal decoding performance based on previous theory in domain generalization} Our decoding analysis achieves 95--99\% accuracy across all modalities and stimulus variables (Table~1). In this section, we provide a theoretical interpretation of this result using the framework of Bayesian risk minimization and domain generalization \citep{Dubois2020, Ruan2021}. In particular, we show why near-optimal decoding implies that our learned embedding retains the stimulus features selectively encoded by V1.

\textbf{Notation.} We use $X$, $Y$, $Z$ to denote input, labels, and the learned embedding, respectively. $p(X)$ (or $p_X$) denotes the probability distribution of $X$, with support $\mathrm{supp}(p_X) = \{x \in \mathcal{X} \mid p_X(x) > 0\}$. We define an encoder as a conditional distribution $p_{Z|X}: \mathcal{Z} \times \mathcal{X} \rightarrow [0,1]$ mapping the input space $\mathcal{X}$ to the representation space $\mathcal{Z}$. We use $f$ to denote a decoder defined on the input $X$ to labels $Y$, and $h$ to denote a decoder defined on the embedding $Z$ to $Y$, with $\mathcal{F}^*$ and $\mathcal{H}^*$ denoting their respective optimal solutions. Following \citep{Dubois2020, Ruan2021}, we use $R(\cdot)$ to denote the Bayesian risk, defined as the infimum over all decoders: $R[Y|X] := \inf_{f} R_f[Y|X] = \inf_{f} \mathbb{E}_{p_{X,Y}}[l(Y,f(X))]$ and $R[Y|Z] := \inf_{h} R_h[Y|Z] = \inf_{h} \mathbb{E}_{p_{Z,Y}}[l(Y,h(Z))]$. For log-loss, this reduces to the conditional entropy: $R[Y|X] = H[Y|X]$. For cross-modality contrastive learning, we define two modalities $M_{\textit{neural}}$ and $M_{\textit{movie}}$. For example, $R_{M_{\textit{neural}}}[Y|Z]$ is the Bayes risk of an embedding $Z$ obtained from neural activity.

We begin with two lemmas from \citet{Dubois2020} on the properties of the Bayesian risk $R[Y|X]$ and $R[Y|Z]$, then follow \citet{Ruan2021} to extend these lemmas to cross-modality setting. We then characterize the optimal representations achievable by cross-modality contrastive learning. We conclude by interpreting our decoding results through this theoretical framework. All proofs can be found in \citet{Dubois2020, Ruan2021}; we include the theorems and lemmas here to make this manuscript self-contained. 

\begin{lemma}[Generalized data processing inequality \cite{Cover2006} for Bayes risk \citep{Xu2020, Dubois2021}]
Let Z-X-Y be a Markov chain of random variables. For any loss function $l$, 
\[
R[Y|X] \leq R[Y|Z]
\]
\end{lemma}

This lemma states that no embedding $Z$ can achieve a lower Bayesian risk than the input $X$ itself. Equality $R[Y|Z] = R[Y|X]$ holds only when $Z$ is an optimal embedding that preserves all information $X$ contains about $Y$. Our encoder defines a Markov chain $Z$-$X$-$Y$: all information $Z$ has about $Y$ must pass through $X$. This constraint leads us to the next lemma.

\begin{lemma}[Equivalence of optimal Bayesian risk and support match]
Let Z-X-Y be a Markov chain of random variables. Then we have
\[
R[Y|Z] = R[Y|X] \quad \Leftrightarrow \quad \forall h^* \in \mathcal{H}^*_{\mathcal{Z}}, \quad \forall(x,z) \in supp(p_{X,Z}), \quad h^*(z) =f^*(x). 
\]
if a) all distributions are discrete; b) there exists a local optimal solution for the loss function $l(\cdot)$
\label{lemma: optimal}
\end{lemma}

This lemma denotes that when an embedding is optimal ($R[Y|Z] = R[Y|X]$), the optimal decoder in the embedding space must agree with the optimal decoder in the input space at every point in their joint support. In other words, decoding $Y$ from $Z$ recovers the same predictions as decoding $Y$ directly from $X$. We now extend this framework to the cross-modality setting.

Cross-modality contrastive learning is a method for tackling the idealized domain generalization problem in representation learning. Here, ``domains'' refer to different modalities---for example, an image of a cat versus the text ``this is a cat.'' The goal is to learn embeddings that capture abstract concepts (e.g., distinguishing cats from dogs) generalizable across modalities. This requires: (a) multiple domains containing decodable information in different formats, and (b) shared labels $Y$ decodable from all domains. In our case, V1 population activity and the natural movie are different domains (we refer to them as modalities in the cross-modality framework), and the shared labels are the stimulus features selectively encoded by V1.

We now extend the Bayesian risk to the idealized domain generalization setting. The following definition is from \citet{Ruan2021}, adapted here to our specific modalities (neural activity and natural movie).
\setcounter{theorem}{0}
\begin{definition}[Bayesian risk for idealized domain generalization between neural activity and natural movie]
Given an encoder $p_{Z|X}$ and a distribution $p_{M_{\textit{neural}},M_{\textit{movie}}}$ on both the neural modality $M_{\textit{neural}}$ and the movie modality $M_{\textit{movie}}$, the idealized domain generalization risk is the expected worst-case risk on the neural modality taken over minimizers in the natural movie modality, i.e.
\[
R_{IDG}[Y|Z] := \mathbb{E}_{p_{M_{\textit{neural}},M_{\textit{movie}}}} \left[\mbox{sup}_{h\in H^*_{Z,M_{\textit{neural}}}}R^{M_{\textit{movie}}}_h[Y|Z]\right]
\]
\end{definition}

\begin{figure}
\centering
         \includegraphics[width=0.8\textwidth]{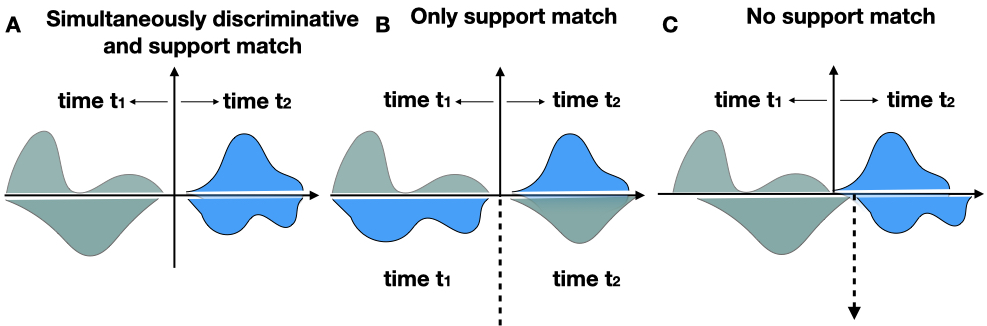}
\caption{A) The upper and lower domains share both their decision boundaries and supports in spite of the difference in the probability distributions (shown as different shapes) ; B) Only support match, no sharing of decision boundaries; C) no support match. }
\label{fig:supportmatch}
\end{figure}

Using this definition, \citep{Ruan2021} introduces the theorem below to characterize the optimal representation. 
\setcounter{theorem}{0}
\begin{theorem}[Optimal representation in cross-modality contrastive learning]
When an optimal encoder $p_{Z^*|X}$ achieves $R[Y|Z^*] = R[Y|X]$, $Z^*$ minimizes 

the risk $R_{IDG}[Y|Z]:= \inf_{h} \mathbb{E}_{p_{M_{\textit{movie}, \textit{neural}}}}[R^{M_{\textit{movie}, \textit{neural}}}_{h}[Y|Z]]$ 

while matching the support of Z across domains, i.e., 
\begin{equation}
R_{IDG}[Y|Z] = R_{IDG}[Y|X] \quad \Leftrightarrow \quad \forall m \in M, \quad supp(p_{Z|m ={neural, movie}}) = supp(p_Z)
\end{equation}
\label{thm:idg}
\end{theorem}
This theorem extends Lemma~\ref{lemma: optimal} to the cross-modality setting. Lemma~\ref{lemma: optimal} requires that optimal decoders agree on the joint support of input and embedding, $\mathrm{supp}(p_{X,Z})$. With multiple modalities, an optimal embedding from one domain must achieve $f^*(X)$ in that domain (Lemma~\ref{lemma: optimal}), and if this embedding is also optimal in another domain, the joint support across domains is their intersection. This is the support matching condition: the support of the neural embedding must agree with that of the movie embedding (\ref{fig:supportmatch}) when those embeddings are optimal. This means that any feature encoded in one modality must also be encoded in all others.

\textbf{Interpretation of our decoding analysis.} Our 95--99\% decoding accuracy across all modalities (Table~1) suggests that our learned embeddings approximate the support matching condition of Theorem~\ref{thm:idg}. The neural embedding therefore captures the common sources of variations across V1 activity and the natural movie. They include the stimulus features selectively encoded by V1.

\subsubsection*{Neural collapse as a candidate geometry in the learned embedding}
\label{appendix:NC}
The Bayesian risk framework above establishes \textit{that} our embedding is near-optimal, but does not describe \textit{what geometric structure} we may observe. Neural collapse \citep{Papyan2020} provides a candidate geometric characterization. It extends previous work on kernel methods \citep{Caponnetto2007, Bordelon2022} and characterizes an intriguing geometry in the last-layer features and classifiers of deep neural networks when training enters the terminal phase (training accuracy plateaus while testing accuracy improves):


\begin{itemize}
    \item[NC1] Variability collapse: The within-class variation of the last-layer features becomes 0. This corresponds to the phenomenon that features collpase to their class means; 
    \item[NC2] All class means collapse to vertices of a simplex equiangular tight frame (ETF) up to scaling
    \item[NC3] Up to scaling, the last-layer classifiers each collapse to the dual of the corresponding class means. 
    \item[NC4] When the network makes an inference on a test example, its decision collapses to simply choosing the class with the closest Euclidean distance between its class mean and the activations of the test example. 
\end{itemize}
\begin{definition}[$K$-Simplex ETF] A standard Simplex ETF is a collection of points in $\mathbb{R}^K$ specified by the columns of 
\begin{equation}
M = \sqrt{\frac{K}{K-1}}\left(I_k - \frac{1}{K} 1_k 1_K^T \right)
\label{ref:etf1}
\end{equation}
where $I_K \in \mathbb{R}^{K\times K}$ is the identity matrix, and $1_K \in \mathbb{R}^K$ is the all ones vector. Alternatively, we may rewrite the Equation \ref{ref:etf1} as: 
\begin{equation}
M^T M = MM^T = \frac{K}{K-1}(I_K - \frac{1}{K} 1_K 1_K^T)
\end{equation}. 
Following the notion introduced in \citep{Papyan2020,Fang2021, Zhu2021}, we consider general Simplex ETF as a collection of points in $\mathbb{R}^d$ specified by the columns of $\sqrt{\frac{K}{K-1}} P \left(I_k - \frac{1}{K} 1_k 1_K^T \right)$, where (i) when $d \geq K$, $P\in \mathbb{R}^{d\times K}$ is an orthonormal matrix, i.e., $P^T P = I_K$, and (ii) when $d = K-1$, P is chosen such that $\left[ P^T \frac{1}{\sqrt{K}}1_K\right]$ is an orthonormal matrix. 
\end{definition}

A large body of work has investigated neural collapse using cross-entropy loss. Recently, \citep{Fang2021} proposed the layer-peeled model, a mathematically tractable surrogate that explains and predicts common patterns in deep neural networks. This model isolates the topmost layer (hence the name) and imposes constraints corresponding to weight decay or normalization applied during training. This top-down approach contrasts with conventional bottom-up analyses that study feature representations starting from the input \citep{Yun2018,Yun2017,Haeffele2015,Baldi1989,Kawaguchi2016,Safran2017,Laurent2017,Zhu2018,Zhou2021,Liang2018,braun2022}. The key insight is that modern overparameterized networks have the capacity to learn arbitrary representations, so last-layer features can be treated as outputs of a universal function approximator.

Here, we include the layer-peeled model for contrastive loss. Given $z_{k,i}$ as the last-layer features for the $i$-th example with label $k$, the layer-peeled model takes the form of
\begin{equation}
\begin{aligned}
\min_{Z} \frac{1}{N}\sum_{k=1}^K \sum_{i=1}^n L_c(z_{k,i}, y_k)\\
\textrm{s.t.} \frac{1}{K} \sum_{k=1}^K \frac{1}{n} \sum_{i=1}^n ||z_{k,i}||^2 \leq c_0\\
\end{aligned}
\end{equation}
where the overall loss function $L_c$ takes the following form (including all training data)
\begin{equation}
\frac{1}{n}\sum_{j=1}^n -\log{\left(\frac{\exp{(z_{k,i}z_{k,j}/\tau)}}{\sum_{k'=1}^K\sum_{l=1}^n \exp{(z_{k,i}z_{k',j}/\tau)}}\right)}
\label{ref:supcon}
\end{equation}

\citep{Fang2021} proved that supervised contrastive loss exhibits neural collapse in its last-layer features. This result motivates us to use neural collapse as a geometric framework. 


\textbf{Summary of our observations.} We observe variability collapse (NC1) during training in both the single-modality and cross-modality phases, and nearest-mean classification (NC4) after training (see \ref{suppfig:NC}). Because contrastive learning loss only uses embedding activation itself, only NC1 and NC4 are relevant for geometry during training. In addition, we evaluated neural collapse on the 80,000 held-out test samples when comparing embedding geometry with and without drift. Together with global smoothness, the learned embedding has a nearest-neighbor local geometry and it is disrupted with drift (Fig. \ref{fig:5A}). We also observe an ETF geometry in the test data after training (see \ref{fig:newNC}) while drift also breaks this ETF structure.

\subsection{Supplementary Results}\label{supp:results}

\subsubsection*{Additional results on the movie stimulus}\label{autocorr2}
%

\begin{figure}
 
     \begin{subfigure}[t]{0.45\textwidth}
     \caption{}
         \centering
         \includegraphics[width=0.9\textwidth]{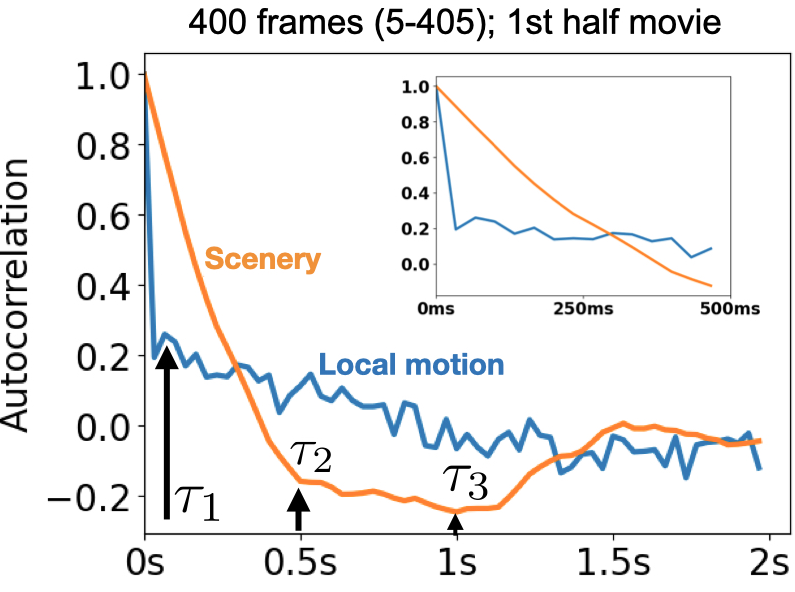}       
         \label{fig:1st}
\end{subfigure}
     \begin{subfigure}[t]{0.45\textwidth}
     \caption{}
         \centering
         \includegraphics[width=0.9\textwidth]{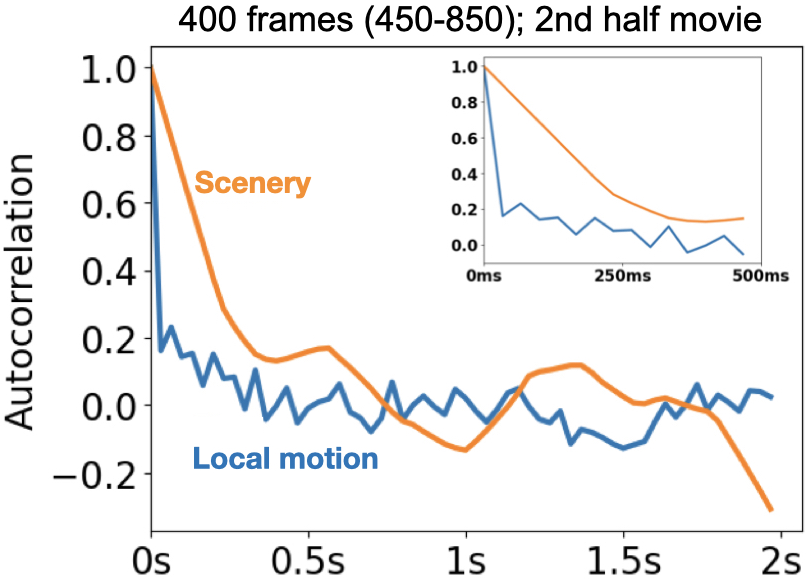}
         \label{fig:2nd}
     \end{subfigure}
\caption{Autocorrelation of the 400 frames in the first and second half of the movie. A) The first 400 frames of the first half start at the fifth frame because this is the center of the first neural activity, with a sampling window around frames 1–10. In B), we first performed hierarchical clustering on both scenery frames and local motion frames. Then we calculated the autocorrelation of the clustering labels. Both steps are the same as A). We observe that the clustering labels for local motion have a fast decay time, while the decay for clusters based on static scenery structure have a long decay, up to 1 s.}
\label{fig:decayscale}
\end{figure}

\begin{figure}[!ht]
\includegraphics[width=0.5\textwidth]{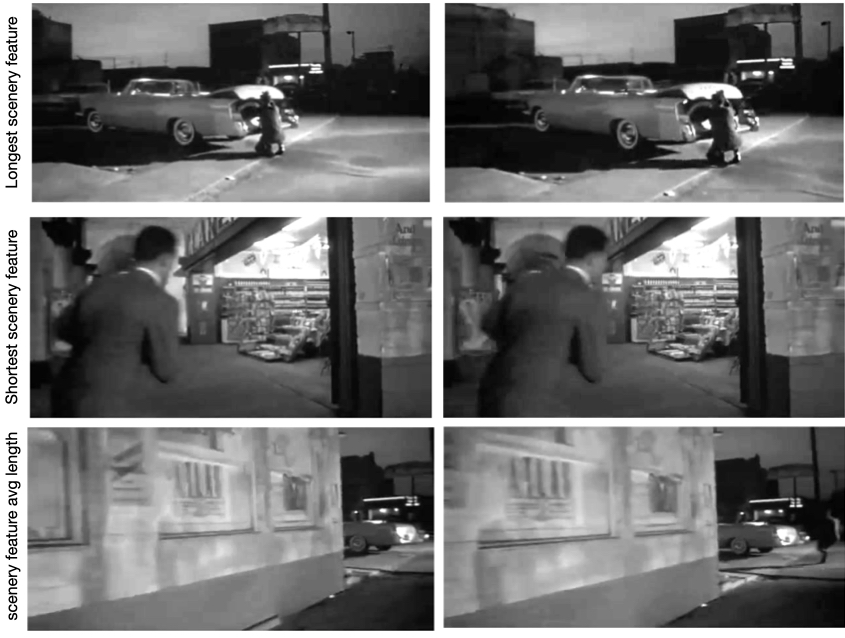}
\caption{\textbf{Scenery feature examples} Hierarchical clustering on scenery features divides 400 movie frames 42 consecutive segments, each containing distinct scenery content as the camera pans through the scene. Here we provide three examples. Because all scenery features are segments of consecutive frames, we show the first and last frame of each segment. The longest scenery feature spans 30 frames (1 second), whereas the shortest spans only 3 frames (100 ms). On average, a distinct scenery feature lasts around 10 frames (330 ms).}
\label{fig:sceneryexample}
\end{figure}

Stimulus features follow a common trend across both halves of the movie. \ref{fig:1st} and \ref{fig:2nd} show that local motion features exhibit fast autocorrelation decay throughout the movie, while scenery features decay much more slowly (also see examples in \ref{fig:sceneryexample}). This pattern holds for both halves, suggesting that the first and second halves contain similar feature statistics. Due to computational costs (Supplementary Materials~\ref{method:pseudo}), we focused on the first half of the movie for learning neural representations.

\begin{figure}
\begin{subfigure}[t]{0.5\textwidth}
\caption{}
         \centering
         \includegraphics[width=0.95\textwidth]{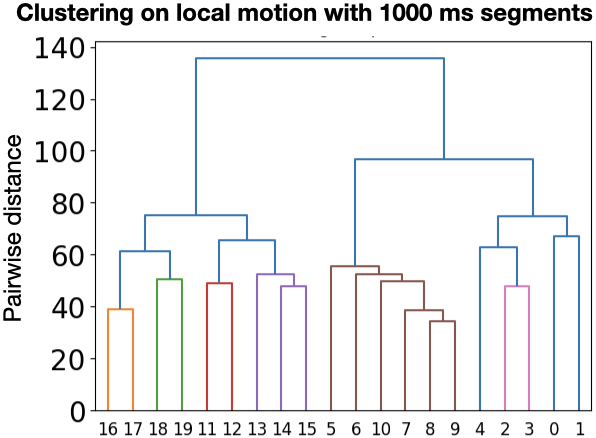}       
         \label{fig:1sflow}
\end{subfigure}
     \begin{subfigure}[t]{0.45\textwidth}
     \caption{}
         \centering
         \includegraphics[width=0.92\textwidth]{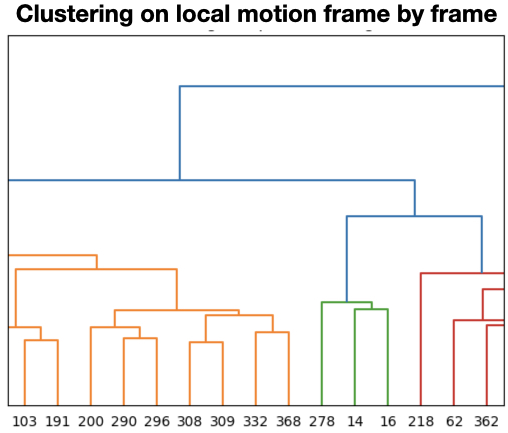}
         \label{fig:30frame}
     \end{subfigure}
\caption{Clustering on coarse 1000 ms segments does not capture fine-grained changes of features in the movie. A) Hierarchical clustering of local motion features using 1-second segments that combine all local motion frames within a specific second, following the same procedure as \ref{fig:allcluster}. The x-axis labels indicate which 1-second segments are being clustered, while colors in the dendrogram show clusters results. For example, all segments from the 5th to 10th seconds belong to the same cluster (shown in brown), indicating that temporal proximity drives clustering rather than feature similarity. B) Hierarchical clustering applied to 399 individual local motion frames (obtained from 400 scenery frames), shown as a detailed view of the clustering in \ref{fig:allcluster}. The x-axis labels show frame numbers and colors represent clustering results. Here, clusters are organized by content similarity rather than temporal order, demonstrating that fine temporal resolution reveals meaningful feature-based groupings that are obscured by coarse temporal binning.}
\label{fig:clustercompare}\end{figure}

Hierarchical clustering produces different results when applied to coarse 1000-ms segments of aggregated local motion frames (following \citep{Xia2021,Sadeh2022}) versus single frames. Coarse-segment clustering (\ref{fig:1sflow}) groups segments primarily by temporal order rather than visual content. In contrast, single-frame clustering (\ref{fig:30frame}) captures content-based similarities. For example, both the 3rd and 10th seconds contain leftward motion—a gentleman moving left and leftward camera movement, respectively—yet temporal clustering places them in separate clusters despite their shared motion characteristics. This demonstrates that averaging over broad timescales (1000 ms) prioritizes temporal proximity over content similarity, failing to preserve information about specific stimulus features.

\subsubsection*{Additional results on the contrastive learning framework}\label{supp:baseline}

In our analyses, we used contrastive learning to learn representations of V1 population activity from which time and stimulus features were linearly decodable, allowing us to compare neural representations of these features between sessions. A key question remains: How difficult is it to decode time in natural scenes directly from neural activity? Given our large neural population size and relatively long sampling window (330 ms), stimulus features might already be explicitly represented and readily decodable from population activity without requiring the expressive power of deep neural networks.

To test this, we examined whether simpler methods could decode time from V1 activity. We trained linear decoders on raw peri-stimulus time histograms (PSTH; 300 units $\times$ 10 time samples = 3,000 dimensions per pseudomouse) and their dimensionality-reduced versions. Table~\ref{tab:naive} shows that linear decoders fail to decode time from raw PSTH, whether from individual pseudomice or combined. Combining both pseudomice with 2048-dimensional PCA improved the decoding performance, but results remained far below contrastive learning. Nonlinear dimensionality reduction methods (t-SNE, Kernel PCA, Isomap) performed even worse ($\sim$0.4--1\% accuracy) and were excluded from the table.

This performance gap reflects a fundamental difference between approaches. Traditional dimensionality reduction reformats the original data space: PCA finds variance-maximizing linear projections, while t-SNE preserves local neighborhood structure. Weakly supervised contrastive learning instead learns an embedding space optimized to maximize agreement between positive pairs (temporally co-occurring neural activity and visual stimuli) while separating negative pairs. This task-specific optimization produces embeddings better suited for decoding behaviorally relevant features than generic dimensionality reduction.

\begin{table}[htp]
    \centering
    \begin{tabular}{|c|c|c|c|c|}\hline
&\textbf{Time} & \textbf{Joint Feature} & \textbf{Scenery} & \textbf{Local motion}\\\hline
p-mouse 1 (raw, 3000D)&8.6\%&11.4\%&16.5\%&11.8\%\\\hline
p-mouse 2 (raw, 3000D)&5.4\%&9.5\%&14.2\%&10.0\%\\\hline
1+2 (raw, 6000D)&8.9\%&15.7\%&24.4\%&15.0\%\\\hline
2048D PCA on (1+2)&13.5\%&22.6\%&35.0\%&19.5\%\\\hline
\end{tabular}
\caption{Decoding performance by naive neural representations (PSTH or PCA). We also used kernel PCA (with nonlinear kernels like RBF and cosine) and TSNE/ISOMAP from the sklearn library for nonlinear dimensionality reduction. Unfortunately, the decoding performance is very low $\sim$0.4\% to 1\%. The chance levels for decoding time, Joint feature, Scenery and Local motion are 0.25\%, 0.5\%, 2.4\% and 1.4\%, respectively}
\label{tab:naive}
\end{table}

During both the pretraining phase and cross-modality training phase, our trained neural backbones can decode time (frame number) with up to 99\% accuracy at single frame resolution (Table \ref{decode:time_allbackbones}). The movie backbones can decode time with high (95.8\%) accuracy as well. We also tested the neural backbone using neural activity from a novel pseudomouse (constructed from a distinct subsets out of the 24 mice available). Note that a) The backbones for scenery and local motion features only receives either scenery or local motion frames as input, whereas the neural backbone retains both scenery and local motion features through neural coding. The better decoding performance from the neural backbone may come from the encoding of scenery and local motion features together. b) the neural backbone only receives supervision from the movie during the cross-modality training phase. 

\begin{table}[]
    \centering
    \begin{tabular}{|c|c|c|c|c|}\hline
   Decode time&$\mbox{Model}_{neural}$&$\mbox{Model}_{scene}$&$\mbox{Model}_{local motion}$&novel pseudomouse (neural)\\\hline
   single-modality  & 99.7\%& 95.8\%&93.8\%&97.7\%\\\hline
   cross-modality  & 99.7\%&96.0\%&95.2\%&98.1\%\\\hline
    \end{tabular}
    \caption{Decoding performance comparison across all backbones in both single-modality and cross-modality phases}
    \label{decode:time_allbackbones}
\end{table}

\begin{figure*}[!ht]
\begin{subfigure}[t]{0.32\textwidth}
\caption{}
         \includegraphics[height=0.75\textwidth, width=0.75\textwidth]{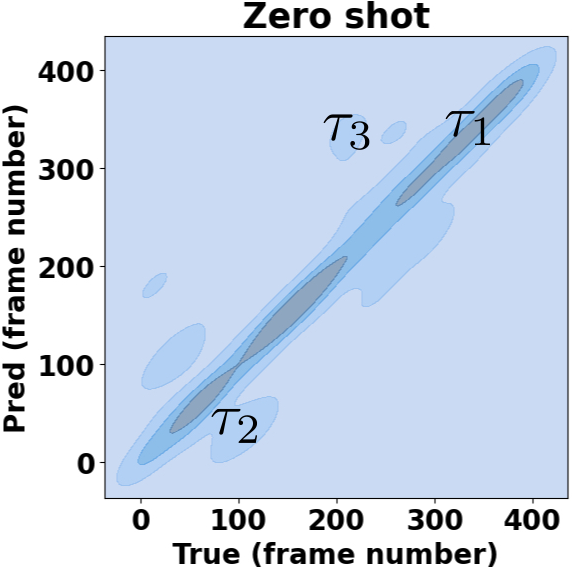}       
         \label{fig:confuseA}
\end{subfigure}
\begin{subfigure}[t]{0.32\textwidth}
\caption{}
         \includegraphics[height=0.78\textwidth, width=0.78\textwidth]{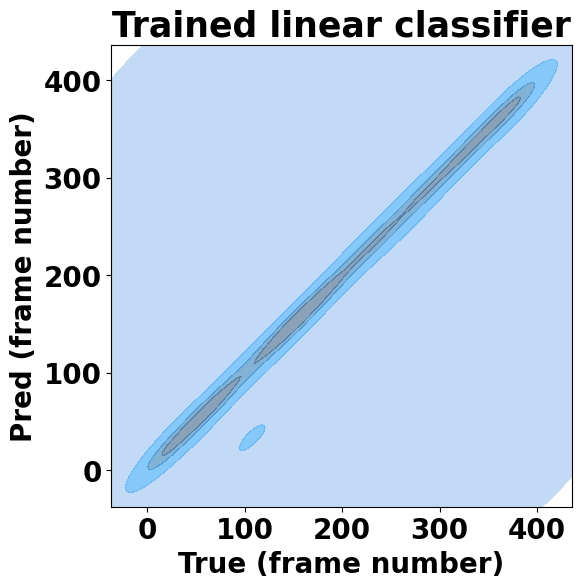}       
         \label{fig:confuseB}
\end{subfigure}
     \begin{subfigure}[t]{0.32\textwidth}
     \caption{}
         \includegraphics[width=0.9\textwidth]{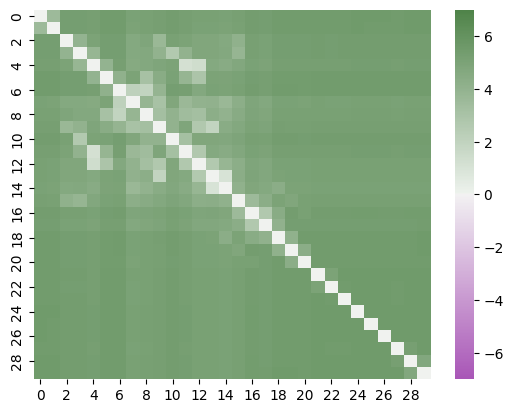}
         \label{fig:pairD}
     \end{subfigure}
\caption{Confusion matrices for decoding time from the neural activity from session 2 passed into the session 1 model; comparison between zero-shot (linear classifier trained on session 1 activations) and a newly trained linear classifier (trained on session 2 activations). A) The kernel density plot of the confusion matrix when we use the linear decoder trained from session 1 to decode natural features from neural activity in session 2. There are three types of decoding errors. Each corresponds to a different temporal scale the main text Results. \ref{sec:movie}. B) The kernel density plot of the confusion matrix for a trained decoder after we train the linear decoder on drifted neural activity. Note that those decoding errors between long-time separated points ($\tau_2$ and $\tau_3$ from A) disappear from B). We hypothesize that the representational drift results in shuffling of time in the natural scenery within their near-time neighbors (as we illustrated in Fig 5C of the main text). C) Pairwise discriminability of 30 frames within a randomly selected 1 second segment. We calculate the discriminability between frame $t$ and $t'$ by first selecting a sub matrix that contains the $t$-th and $t'$-th columns and rows from the confusion matrix in A) and then obtain the hit rate and miss rate from this sub matrix. Then we compute the discriminability $d'$ as the difference between the inverse of standard normal distribution using the hit rate and the miss rate.}
\label{fig:driftdecode}
\end{figure*}

After representational drift, the performance of linear decoding degrades similarly in both pseudomice (Table~\ref{tab:perf}). This suggests that the decay of decoding performance is a stable, reliable estimate of the magnitude of how drift changes the encoding of visual features. Examining the confusion matrices (\ref{fig:driftdecode}B), we find that retraining a linear classifier on drifted activity corrects \enquote{side-banded} errors and narrows confusion along the main diagonal—that is, retraining mostly corrects errors at broad timescales. This pattern suggests that drift shuffles the encoding of stimulus features among their temporal neighbors (Fig.~\ref{fig:5C}), making fine-scale linear decoding challenging even after retraining.

\begin{table}
    \centering
    \begin{tabular}{|c|c|c|c|c|}\hline
    Session 1 to 2 (p-mouse 1) & Time & All features & scenery & Flow\\\hline
    zero-shot &45.6\%&44.6\%&54.4\%&47.1\%\\\hline
    trained linear classifier&85.7\%&71.5\%&77.2\%&69.3\%\\\hline
    Session 2 to 1 (p-mouse 1)& Time & All features & scenery & Flow\\\hline
    zero-shot &47.8\%&47.4\%&56.0\%&47.0\%\\\hline
    trained linear classifier&86.4\%&68.6\%&76.3\%&66.5\%\\\hline

    Session 1 to 2 (p-mouse 2)& Time & All features & scenery & Flow\\\hline
    zero-shot &47.3\%&48.4&58.1\%&47.1\%\\\hline
    trained classifier &78.9\%&64.5\%&75.4\%&64.4\%\\\hline 
    Session 2 to 1 (p-mouse 2)& Time & All features & scenery & Flow\\\hline
    zero-shot &48.8\%&49.4&58.3\%&48.5\%\\\hline
    trained classifer &79.0\%&65.9&74.8\%&63.6\%\\\hline
    
    \end{tabular}
\caption{Decoding performance of individual pseudomouse. All zero-shot performances are reported using linear decoder trained on the corresponding neural activity without drift (at 99\% decoding performance) and tested on the neural activity with drift (no retraining). We observe that the representational drift impact decoding symmetrically, i.e., models trained on session 2 show a similar 40\% degradation when they are used to decode neural activity from session 1. It is possible that in session 2, animals may have different internal states than session 1 because the movie becomes familiar. This symmetry suggests that the neural representation successfully emphasizes stimulus-relevant features over internal state. The high agreement between both pseudomice and between sessions suggests that our quantification of representational drift has sufficient sampling.}
\label{tab:perf}
\end{table}

\subsubsection*{Additional results on the embedding geometry}\label{sec:optimalgeometry}

\begin{figure}
     \begin{subfigure}[t]{0.45\textwidth}
     \caption{}
         \centering
         \includegraphics[width=0.9\textwidth]{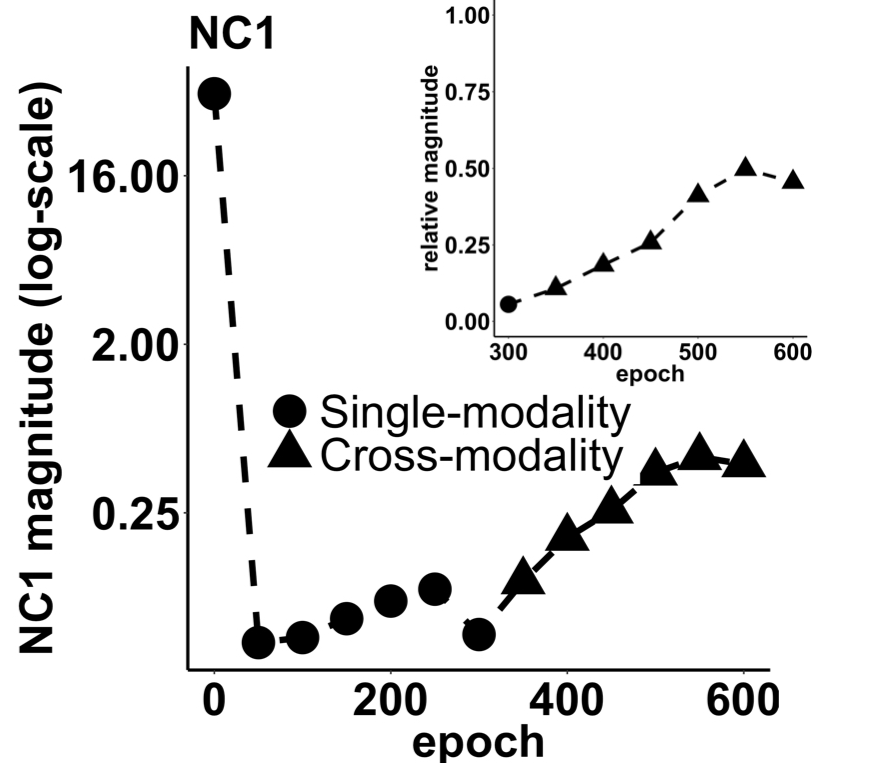}       
         \label{fig:nc1}
\end{subfigure}
     \begin{subfigure}[t]{0.40\textwidth}
     \caption{}
         \centering
         \includegraphics[width=0.9\textwidth]{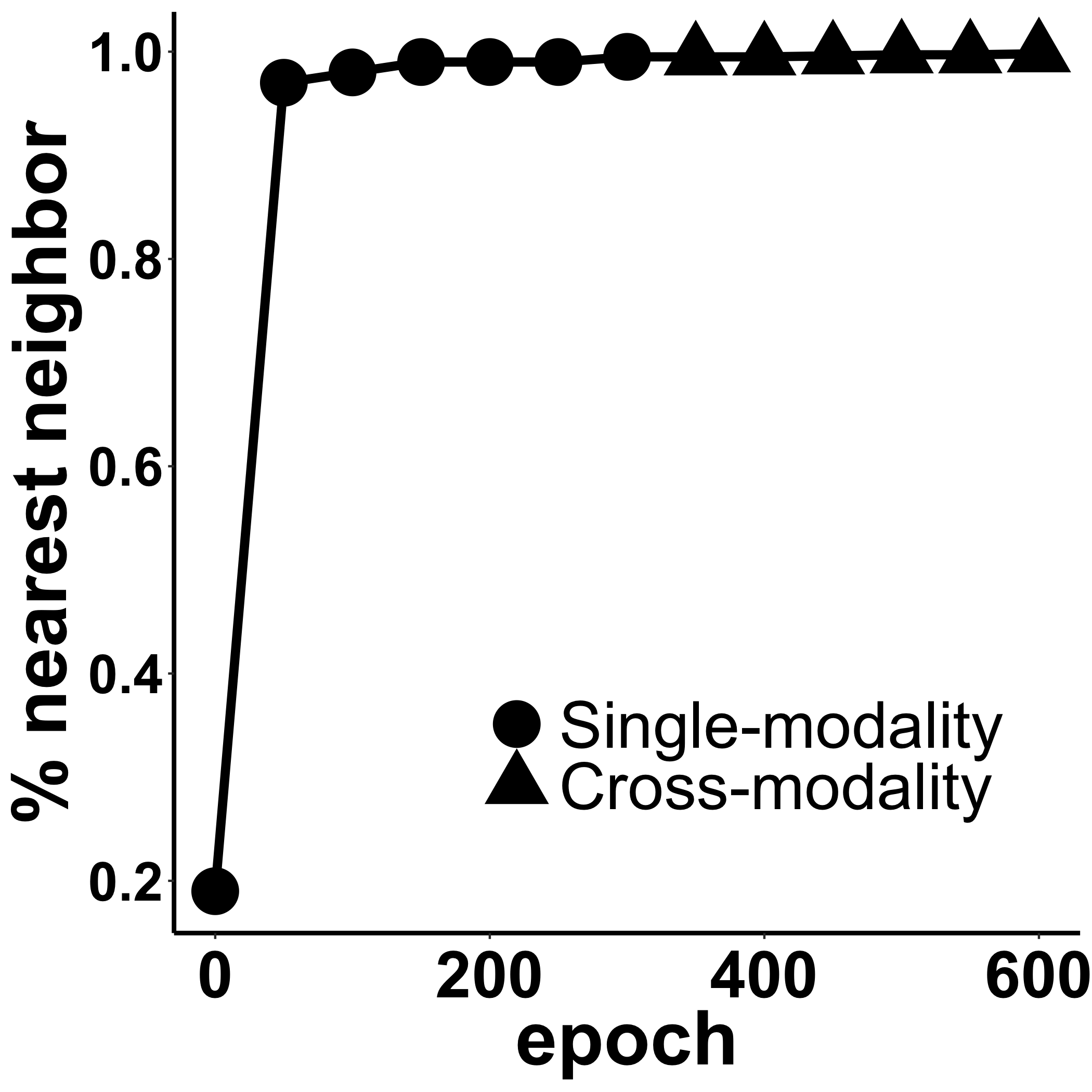}
         \label{fig:nc4}
     \end{subfigure}
\caption{Neural collapse geometry during training. A) The variability of activations for neural activity centered at any time $t$(frame number) approaches zero. This is the variability collapse observed for the first time in \cite{Papyan2020}. Similar to \citep{Papyan2020, Zhu2021}, we measure the variability collapse as the relative magnitude of within-cluster covariance versus between-class covariance. If this relative magnitude is less than 1, it indicates that the activations within individual frames coalesce into distinct groups. During the single-modal contrastive learning phase, we observe that the variability becomes significantly less than 1 ($\sim$ 0.1). In the cross-modality phase, the variability remains low (significantly less than 1; see inset) but increases slightly. Note that the collapse of variability was only demonstrated under full supervision. Here we may be seeing that weak supervision resulting from co-occurrence could result in different training dynamics. B) Cross-modal contrastive learning preserves the neural collapse behavior that emerges from single-modal contrastive learning. This behavior corresponds to simplifying the classification of samples by finding the nearest mean activation (dubbed "NC4"). Here, we demonstrate the proportion of samples (\% of the hold-out test set) that can be correctly classified by locating the nearest class mean. We observe that the performance of cross-modal and single-modal contrastive learning is comparable.}
\label{suppfig:NC}
\end{figure}

We examined whether this performance is accompanied by geometrical changes in the embeddings associated with so-called neural collapse. Neural collapse refers to a phenomenon observed in deep neural networks during terminal-phase training, where within-class variability becomes minimal and class means align to form a simplex equiangular tight frame (ETF) (Supplementary Materials~\ref{appendix:NC}). While neural collapse has been characterized primarily in fully supervised settings \citep{Papyan2020}, we ask whether we can observe similar geometry in our embedding learned from weakly supervised contrastive learning. Two properties of neural collapse are relevant. First, variability collapse (\enquote{NC1}): feature activation variability for each class approaches zero, causing activations to form distinct clusters. Second, nearest-mean classification (\enquote{NC4}): classification reduces to finding the class with the nearest mean activation. As training progresses, our embedding exhibits decreasing variability (\ref{fig:nc1}) and nearest-mean behavior (\ref{fig:nc4}), indicating that activations at different times form disjoint clusters. We observe that variability does not collapse completely to zero on the test set (\ref{fig:nc1}), yet remains sufficiently low. At the same time, both single-modality and cross-modality models achieve 97\% accuracy on a novel pseudomouse (Table~1), indicating that this small change in NC1 does not impact decoding performance. This non-zero variability while maintaining discriminability may reflect differences between weakly supervised contrastive learning and fully supervised classification, where neural collapse was originally characterized. Whether these differences affect generalization remains an open question. 

\begin{figure}
    \centering
         \includegraphics[width=0.45\textwidth]{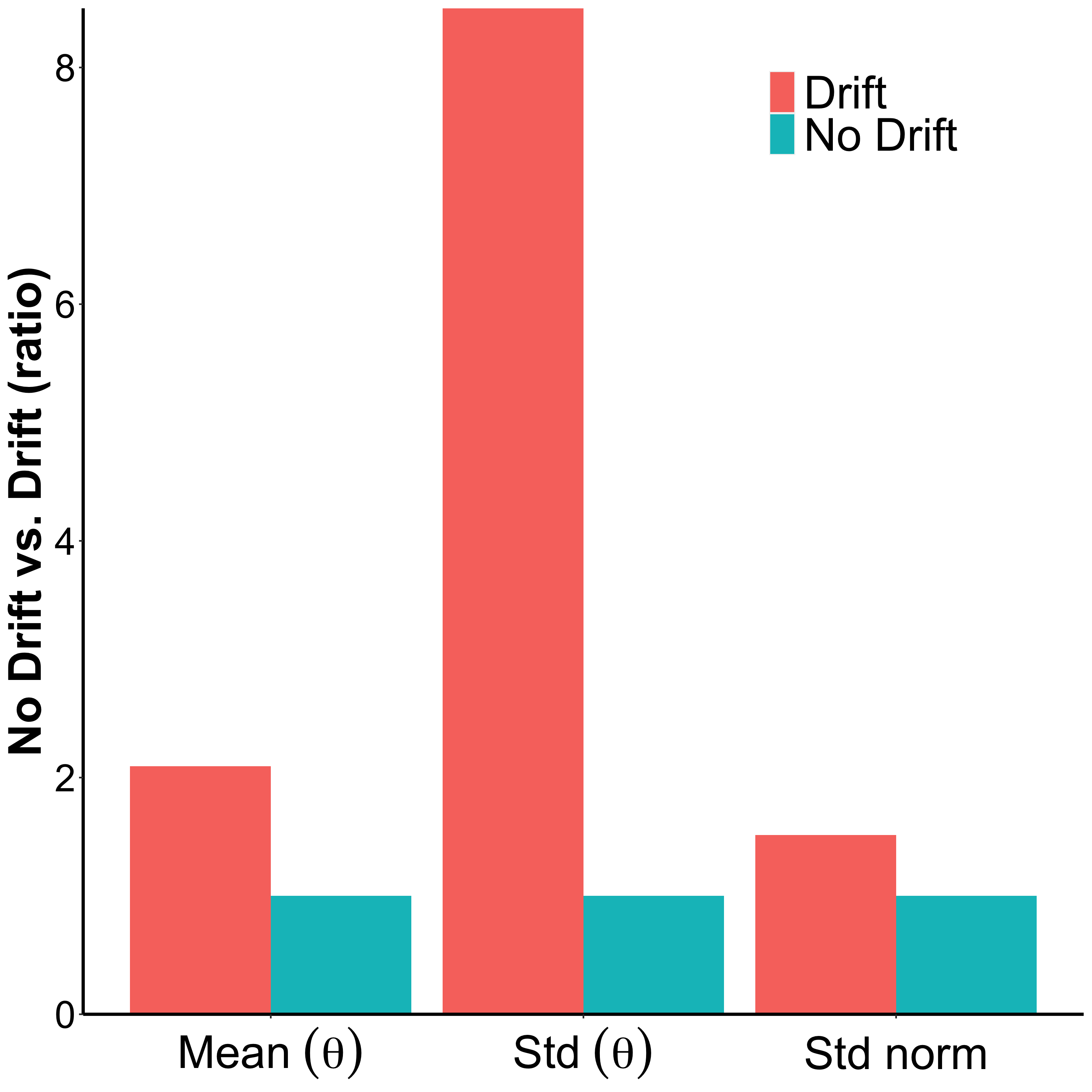}
    \caption{\textbf{Representational drift alters the geometry of neural embedding clusters.} Individual clusters represent all neural samples from the same temporal window centered at time $t$. We quantify geometric changes using two metrics: angles $\theta$ between all pairs of cluster centroids, and cluster norms defined as the mean pairwise distance between all samples within the same cluster. Without drift, both metrics remain low, indicating consistent inter-cluster arrangements and tight cluster organization. Representational drift increases both the mean and standard deviation of $\theta$, as well as cluster norms, demonstrating that drift simultaneously disrupts cluster positioning and increases within-cluster spread.}
    \label{fig:newNC}
\end{figure}

In \ref{fig:newNC}, we provide evidence for the emergence of simplex ETF geometry without drift and demonstrate how representational drift perturbs this structure. The simplex ETF framework predicts that when samples from the same time window $t$ form distinct clusters separate from samples at any other time $t'$, the angles between cluster centroids become equal after the contrastive learning model achieves optimal classification performance. By defining $\theta$ as angles between all cluster centroids and plotting both $mean(\theta)$ and $std(\theta)$, we demonstrate that such a simplex ETF emerges in our learned embedding (teal bars in \ref{fig:newNC}). In contrast, representational drift disrupts this simplex ETF geometry by substantially increasing both the mean and standard deviation of $\theta$ (coral bars in \ref{fig:newNC}). Simultaneously, drift increases cluster norms, disrupting the variability collapse that also characterizes optimal embedding geometry. We illustrate this disrupted geometry in Fig. \ref{fig:5C}.

\begin{figure}
    \centering
         \includegraphics[width=0.45\textwidth]{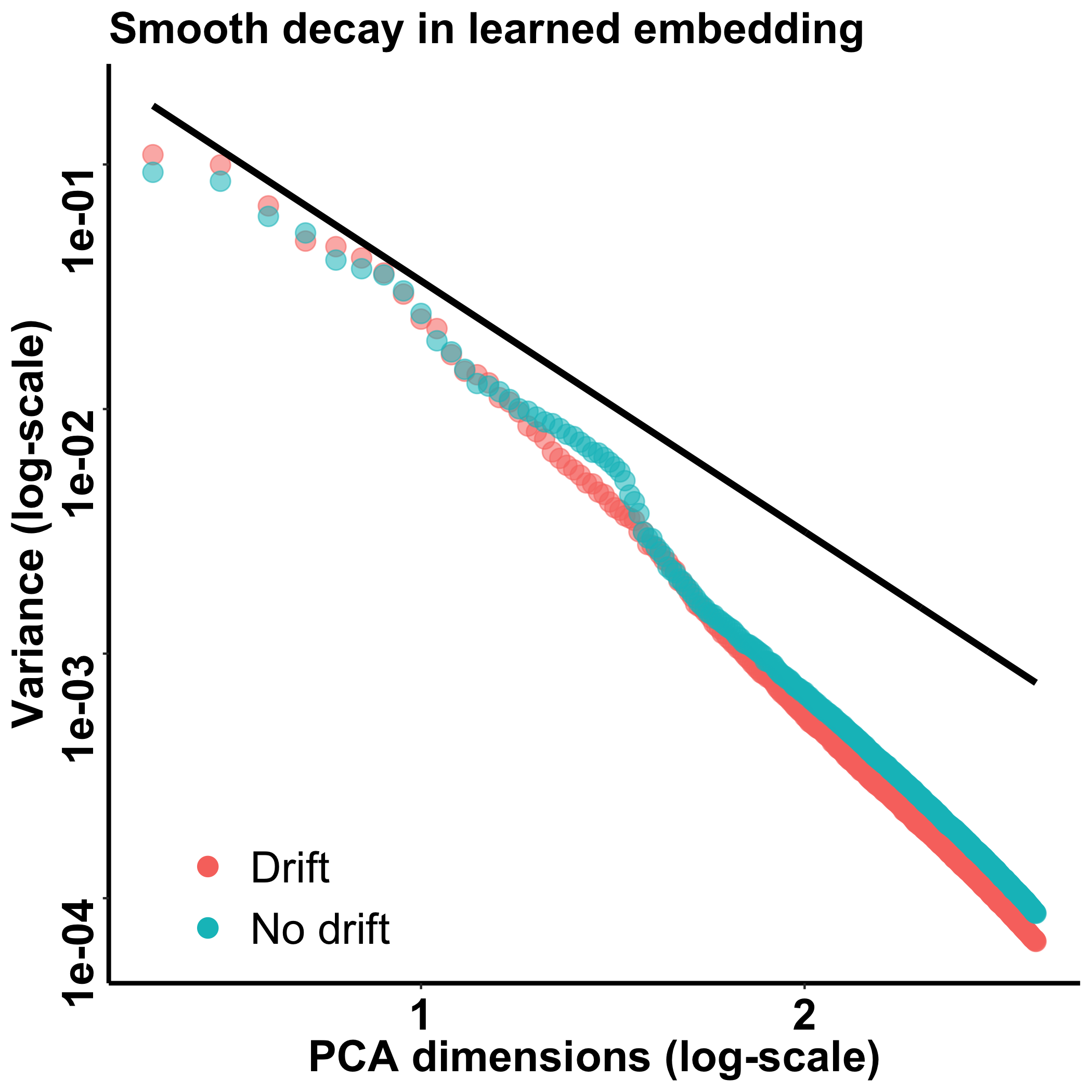}
    \caption{The global geometry of the learned embedding: Principal components (PC) with or without drift show fast decays. These decays are faster than the threshold $-1-2/d$ ($d$ as the stimulus dimension), suggesting that both embedding are smooth manifolds. Two neural encodings of close stimulus features are nearby in the embedding space because of this smoothness. }
         \label{fig:pcadecay}
\end{figure}

We examine whether representational drift disrupts the global geometry of our learned embedding. If drift impairs global structure, this could explain the increased decoding errors. To test this, we analyze \enquote{coding smoothness} introduced by \citet{Stringer2019a}, which characterizes whether similar stimulus features map to nearby locations in representation space. The diagnostic for smoothness is the decay rate of principal component variance: if this decay is faster than a threshold of $(-1-2/d)$ for a $d$-dimensional stimulus, the representation is smooth and forms a differentiable manifold. \citet{Stringer2019a} showed that neural activity satisfies this condition when encoding static natural images. \ref{fig:pcadecay} shows that the variance of principal components in our learned embedding decays much faster than the threshold (with $d=400$), indicating global smoothness. This smoothness is also preserved after representational drift. This result suggests that drift does not disrupt global manifold structure; instead, the observed decoding errors must arise from perturbations to local geometry (as analyzed in Results.~\ref{sec:driftgeometry}).

\section*{Acknowledgement}
This work was supported by the National Science Foundation through the Physics Frontier Center for Living Systems (PHY-2317138) and the NSF-Simons National Institute for Theory and Mathematics in Biology (NITMB), awards NSF DMS-2235451 and Simons Foundation MP-TMPS-00005320. This work was also supported by the National Science Foundation Graduate Research Fellowship Program (DGE-1746045) to EAdL. 

\bibliographystyle{plainnat}
\bibliography{covariate}

\begin{thebibliography}{91}
\providecommand{\natexlab}[1]{#1}
\providecommand{\url}[1]{\texttt{#1}}
\expandafter\ifx\csname urlstyle\endcsname\relax
  \providecommand{\doi}[1]{doi: #1}\else
  \providecommand{\doi}{doi: \begingroup \urlstyle{rm}\Url}\fi

\bibitem[Aitken et~al.(2022)Aitken, Garrett, Olsen, and Mihalas]{Aitken2022}
Kyle Aitken, Marina Garrett, Shawn Olsen, and Stefan Mihalas.
\newblock The geometry of representational drift in natural and artificial neural networks.
\newblock \emph{PLoS Comput. Biol.}, 18\penalty0 (11):\penalty0 e1010716, November 2022.

\bibitem[Attardo et~al.(2015)Attardo, Fitzgerald, and Schnitzer]{Attardo2015}
Alessio Attardo, James~E Fitzgerald, and Mark~J Schnitzer.
\newblock Impermanence of dendritic spines in live adult {CA1} hippocampus.
\newblock \emph{Nature}, 523\penalty0 (7562):\penalty0 592--596, July 2015.

\bibitem[Baldi and Hornik(1989)]{Baldi1989}
Pierre Baldi and Kurt Hornik.
\newblock Neural networks and principal component analysis: Learning from examples without local minima.
\newblock \emph{Neural Networks}, 2\penalty0 (1):\penalty0 53--58, jan 1989.
\newblock \doi{10.1016/0893-6080(89)90014-2}.

\bibitem[Barra et~al.(2020)Barra, Carta, Corriga, Podda, and Recupero]{Barra2020}
Silvio Barra, Salvatore~Mario Carta, Andrea Corriga, Alessandro~Sebastian Podda, and Diego~Reforgiato Recupero.
\newblock Deep learning and time series-to-image encoding for financial forecasting.
\newblock \emph{IEEE/CAA Journal of Automatica Sinica}, 7\penalty0 (3):\penalty0 683–692, May 2020.
\newblock ISSN 2329-9274.
\newblock \doi{10.1109/jas.2020.1003132}.
\newblock URL \url{http://dx.doi.org/10.1109/JAS.2020.1003132}.

\bibitem[Bordelon and Pehlevan(2022)]{Bordelon2022}
Blake Bordelon and Cengiz Pehlevan.
\newblock Population codes enable learning from few examples by shaping inductive bias.
\newblock \emph{Elife}, 11\penalty0 (e78606), December 2022.

\bibitem[brain map.org(2019)]{Allen2019}
brain map.org.
\newblock Allen brain observatory – neuropixels visual coding.
\newblock Technical report, 2019.
\newblock URL \url{https://portal.brain-map.org/explore/circuits/visual-coding-neuropixels}.

\bibitem[Braun et~al.(2022)Braun, Domin{\'e}, Fitzgerald, and Saxe]{braun2022}
Lukas Braun, Cl{\'e}mentine Carla~Juliette Domin{\'e}, James~E Fitzgerald, and Andrew~M Saxe.
\newblock Exact learning dynamics of deep linear networks with prior knowledge.
\newblock In Alice~H. Oh, Alekh Agarwal, Danielle Belgrave, and Kyunghyun Cho, editors, \emph{Advances in Neural Information Processing Systems}, 2022.
\newblock URL \url{https://openreview.net/forum?id=lJx2vng-KiC}.

\bibitem[Brenner et~al.(2000)Brenner, Strong, Koberle, Bialek, and de~Ruyter~van Steveninck]{Brenner2000}
N~Brenner, S~P Strong, R~Koberle, W~Bialek, and R~R de~Ruyter~van Steveninck.
\newblock Synergy in a neural code.
\newblock \emph{Neural Comput.}, 12\penalty0 (7):\penalty0 1531--1552, July 2000.

\bibitem[Buzsáki(2019)]{Buzs_ki_2019}
György Buzsáki.
\newblock \emph{The Brain from Inside Out}.
\newblock Oxford University PressNew York, June 2019.
\newblock ISBN 9780190905415.
\newblock \doi{10.1093/oso/9780190905385.001.0001}.

\bibitem[Caponnetto and Vito(2007)]{Caponnetto2007}
A.~Caponnetto and E.~Vito.
\newblock Optimal rates for the regularized least-squares algorithm.
\newblock \emph{Found. Comput. Math.}, 7\penalty0 (3):\penalty0 331–368, July 2007.
\newblock ISSN 1615-3375.

\bibitem[Chen et~al.(2020{\natexlab{a}})Chen, Kornblith, Norouzi, and Hinton]{Chen2020}
Ting Chen, Simon Kornblith, Mohammad Norouzi, and Geoffrey Hinton.
\newblock A simple framework for contrastive learning of visual representations.
\newblock In Hal~Daumé III and Aarti Singh, editors, \emph{Proceedings of the 37th International Conference on Machine Learning}, volume 119 of \emph{Proceedings of Machine Learning Research}, pages 1597--1607. PMLR, 13--18 Jul 2020{\natexlab{a}}.
\newblock URL \url{https://proceedings.mlr.press/v119/chen20j.html}.

\bibitem[Chen et~al.(2020{\natexlab{b}})Chen, Kornblith, Swersky, Norouzi, and Hinton]{Chen2020b}
Ting Chen, Simon Kornblith, Kevin Swersky, Mohammad Norouzi, and Geoffrey Hinton.
\newblock Big self-supervised models are strong semi-supervised learners.
\newblock In \emph{Proceedings of the 34th International Conference on Neural Information Processing Systems}, NIPS '20, Red Hook, NY, USA, 2020{\natexlab{b}}. Curran Associates Inc.
\newblock ISBN 9781713829546.

\bibitem[Cho et~al.(2015)Cho, Lee, Shin, Choy, and Do]{cho2015much}
Junghwan Cho, Kyewook Lee, Ellie Shin, Giles Choy, and Synho Do.
\newblock How much data is needed to train a medical image deep learning system to achieve necessary high accuracy?
\newblock \emph{arXiv preprint arXiv:1511.06348}, 2015.

\bibitem[Climer et~al.(2025)Climer, Davoudi, Oh, and Dombeck]{Climer2025}
Jason~R Climer, Heydar Davoudi, Jun~Young Oh, and Daniel~A Dombeck.
\newblock Hippocampal representations drift in stable multisensory environments.
\newblock \emph{Nature}, 645\penalty0 (8080):\penalty0 457--465, September 2025.

\bibitem[Cover and Thomas(2006)]{Cover2006}
Thomas~M. Cover and Joy~A. Thomas.
\newblock \emph{Elements of Information Theory 2nd Edition (Wiley Series in Telecommunications and Signal Processing)}.
\newblock Wiley-Interscience, July 2006.
\newblock ISBN 0471241954.

\bibitem[De~Pasquale et~al.(2014)De~Pasquale, Beckhauser, Hernandes, and Giorgetti~Britto]{De_Pasquale2014}
Roberto De~Pasquale, Thiago~F Beckhauser, Marina~Sorrentino Hernandes, and Luiz~R Giorgetti~Britto.
\newblock {LTP} and {LTD} in the visual cortex require the activation of {NOX2}.
\newblock \emph{J. Neurosci.}, 34\penalty0 (38):\penalty0 12778--12787, September 2014.

\bibitem[Deitch et~al.(2021)Deitch, Rubin, and Ziv]{Deitch2021}
Daniel Deitch, Alon Rubin, and Yaniv Ziv.
\newblock Representational drift in the mouse visual cortex.
\newblock \emph{Current Biology}, 31\penalty0 (19):\penalty0 4327--4339.e6, oct 2021.
\newblock \doi{10.1016/j.cub.2021.07.062}.

\bibitem[Delamare et~al.(2024)Delamare, Zaki, Cai, and Clopath]{Delamare2024}
Geoffroy Delamare, Yosif Zaki, Denise~J Cai, and Claudia Clopath.
\newblock Drift of neural ensembles driven by slow fluctuations of intrinsic excitability.
\newblock \emph{Elife}, 12\penalty0 (RP88053), May 2024.

\bibitem[Devalle et~al.(2025)Devalle, Zou, Cecchini, and Roxin]{Devalle2025}
Federico Devalle, Licheng Zou, Gloria Cecchini, and Alex Roxin.
\newblock Representational drift as the consequence of ongoing memory storage.
\newblock \emph{Sci. Rep.}, 15\penalty0 (1):\penalty0 27746, July 2025.

\bibitem[Driscoll et~al.(2017)Driscoll, Pettit, Minderer, Chettih, and Harvey]{Driscoll2017}
Laura~N. Driscoll, Noah~L. Pettit, Matthias Minderer, Selmaan~N. Chettih, and Christopher~D. Harvey.
\newblock Dynamic reorganization of neuronal activity patterns in parietal cortex.
\newblock \emph{Cell}, 170\penalty0 (5):\penalty0 986--999.e16, aug 2017.
\newblock \doi{10.1016/j.cell.2017.07.021}.

\bibitem[Driscoll et~al.(2022)Driscoll, Duncker, and Harvey]{Driscoll2022}
Laura~N. Driscoll, Lea Duncker, and Christopher~D. Harvey.
\newblock Representational drift: Emerging theories for continual learning and experimental future directions.
\newblock \emph{Current opinion in neurobiology}, 76:\penalty0 102609, October 2022.
\newblock ISSN 1873-6882.
\newblock \doi{10.1016/j.conb.2022.102609}.

\bibitem[Dubois et~al.(2020)Dubois, Kiela, Schwab, and Vedantam]{Dubois2020}
Yann Dubois, Douwe Kiela, David~J Schwab, and Ramakrishna Vedantam.
\newblock Learning optimal representations with the decodable information bottleneck.
\newblock In H.~Larochelle, M.~Ranzato, R.~Hadsell, M.F. Balcan, and H.~Lin, editors, \emph{Advances in Neural Information Processing Systems}, volume~33, pages 18674--18690. Curran Associates, Inc., 2020.
\newblock URL \url{https://proceedings.neurips.cc/paper_files/paper/2020/file/d8ea5f53c1b1eb087ac2e356253395d8-Paper.pdf}.

\bibitem[Dubois et~al.(2021)Dubois, Bloem-Reddy, Ullrich, and Maddison]{Dubois2021}
Yann Dubois, Benjamin Bloem-Reddy, Karen Ullrich, and Chris~J. Maddison.
\newblock Lossy compression for lossless prediction.
\newblock In A.~Beygelzimer, Y.~Dauphin, P.~Liang, and J.~Wortman Vaughan, editors, \emph{Advances in Neural Information Processing Systems}, 2021.
\newblock URL \url{https://openreview.net/forum?id=wZrOOO9XBn}.

\bibitem[Fan et~al.(2023)Fan, Wen, and Lai]{Fan2023}
Junjun Fan, Jiajun Wen, and Zhihui Lai.
\newblock Myoelectric pattern recognition using gramian angular field and convolutional neural networks for muscle-computer interface.
\newblock \emph{Sensors (Basel)}, 23\penalty0 (5):\penalty0 2715, March 2023.

\bibitem[Fang et~al.(2021)Fang, He, Long, and Su]{Fang2021}
Cong Fang, Hangfeng He, Qi~Long, and Weijie~J. Su.
\newblock Exploring deep neural networks via layer-peeled model: Minority collapse in imbalanced training.
\newblock \emph{Proceedings of the National Academy of Sciences of the United States of America}, 118, October 2021.
\newblock ISSN 1091-6490.
\newblock \doi{10.1073/pnas.2103091118}.

\bibitem[Feldman(2012)]{Feldman2012}
Daniel~E Feldman.
\newblock The spike-timing dependence of plasticity.
\newblock \emph{Neuron}, 75\penalty0 (4):\penalty0 556--571, August 2012.

\bibitem[Gale and Murphy(2016)]{Gale2016}
S~D Gale and G~J Murphy.
\newblock Active dendritic properties and local inhibitory input enable selectivity for object motion in mouse superior colliculus neurons.
\newblock \emph{J. Neurosci.}, 36\penalty0 (35):\penalty0 9111--9123, August 2016.

\bibitem[Haeffele and Vidal(2015)]{Haeffele2015}
Benjamin~D. Haeffele and Ren{\'{e}} Vidal.
\newblock Global optimality in tensor factorization, deep learning, and beyond.
\newblock \emph{CoRR}, abs/1506.07540, 2015.
\newblock URL \url{http://arxiv.org/abs/1506.07540}.

\bibitem[Haimerl and Machens(2025)]{Haimerl2025}
Caroline Haimerl and Christian Machens.
\newblock Representational drift without synaptic plasticity.
\newblock \emph{bioRxiv}, 2025.
\newblock \doi{10.1101/2025.07.23.666352}.
\newblock URL \url{https://www.biorxiv.org/content/early/2025/07/29/2025.07.23.666352}.

\bibitem[Hengen et~al.(2013)Hengen, Lambo, Van~Hooser, Katz, and Turrigiano]{Hengen2013}
Keith~B Hengen, Mary~E Lambo, Stephen~D Van~Hooser, Donald~B Katz, and Gina~G Turrigiano.
\newblock Firing rate homeostasis in visual cortex of freely behaving rodents.
\newblock \emph{Neuron}, 80\penalty0 (2):\penalty0 335--342, October 2013.

\bibitem[Jiang et~al.(2010)Jiang, Huang, de~Pasquale, Millman, Song, Lee, Tsumoto, and Kirkwood]{Jiang2010}
Bin Jiang, Shiyong Huang, Roberto de~Pasquale, Daniel Millman, Lihua Song, Hey-Kyoung Lee, Tadaharu Tsumoto, and Alfredo Kirkwood.
\newblock The maturation of {GABAergic} transmission in visual cortex requires endocannabinoid-mediated {LTD} of inhibitory inputs during a critical period.
\newblock \emph{Neuron}, 66\penalty0 (2):\penalty0 248--259, April 2010.

\bibitem[Kawaguchi(2016)]{Kawaguchi2016}
Kenji Kawaguchi.
\newblock Deep learning without poor local minima, 2016.
\newblock URL \url{https://arxiv.org/abs/1605.07110}.

\bibitem[Khosla et~al.(2020)Khosla, Teterwak, Wang, Sarna, Tian, Isola, Maschinot, Liu, and Krishnan]{Khosla2020}
Prannay Khosla, Piotr Teterwak, Chen Wang, Aaron Sarna, Yonglong Tian, Phillip Isola, Aaron Maschinot, Ce~Liu, and Dilip Krishnan.
\newblock Supervised contrastive learning.
\newblock In \emph{Proceedings of the 34th International Conference on Neural Information Processing Systems}, NIPS '20, Red Hook, NY, USA, 2020. Curran Associates Inc.
\newblock ISBN 9781713829546.

\bibitem[K{\"u}hn et~al.(2025)K{\"u}hn, Li, Baimacheva, Zimmer, Reinhard, Bonin, and Farrow]{Kuhn2025}
Norma~K K{\"u}hn, Chen Li, Natalia Baimacheva, Janne Zimmer, Katja Reinhard, Vincent Bonin, and Karl Farrow.
\newblock Dendritic architecture enables de novo computation of salient motion in the superior colliculus.
\newblock \emph{Curr. Biol.}, 35\penalty0 (16):\penalty0 3799--3811.e8, August 2025.

\bibitem[Laurent and von Brecht(2017)]{Laurent2017}
Thomas Laurent and James von Brecht.
\newblock Deep linear neural networks with arbitrary loss: All local minima are global.
\newblock \emph{CoRR}, abs/1712.01473, 2017.
\newblock URL \url{http://arxiv.org/abs/1712.01473}.

\bibitem[Liang et~al.(2015)Liang, Xiong, Zingg, Ji, Zhang, and Tao]{Liang2015}
Feixue Liang, Xiaorui~R. Xiong, Brian Zingg, Xu-ying Ji, Li~I. Zhang, and Huizhong~W. Tao.
\newblock Sensory cortical control of a visually induced arrest behavior via corticotectal projections.
\newblock \emph{Neuron}, 86:\penalty0 755--767, May 2015.
\newblock ISSN 1097-4199.
\newblock \doi{10.1016/j.neuron.2015.03.048}.

\bibitem[LIANG et~al.(2018)LIANG, Sun, Lee, and Srikant]{Liang2018}
SHIYU LIANG, Ruoyu Sun, Jason~D Lee, and R.~Srikant.
\newblock Adding one neuron can eliminate all bad local minima.
\newblock In S.~Bengio, H.~Wallach, H.~Larochelle, K.~Grauman, N.~Cesa-Bianchi, and R.~Garnett, editors, \emph{Advances in Neural Information Processing Systems}, volume~31. Curran Associates, Inc., 2018.
\newblock URL \url{https://proceedings.neurips.cc/paper_files/paper/2018/file/a012869311d64a44b5a0d567cd20de04-Paper.pdf}.

\bibitem[Ma et~al.(2019)Ma, Turrigiano, Wessel, and Hengen]{Ma2019}
Zhengyu Ma, Gina~G Turrigiano, Ralf Wessel, and Keith~B Hengen.
\newblock Cortical circuit dynamics are homeostatically tuned to criticality in vivo.
\newblock \emph{Neuron}, 104\penalty0 (4):\penalty0 655--664.e4, November 2019.

\bibitem[Manns et~al.(2007)Manns, Howard, and Eichenbaum]{Manns2007}
Joseph~R. Manns, Marc~W. Howard, and Howard Eichenbaum.
\newblock Gradual changes in hippocampal activity support remembering the order of events.
\newblock \emph{Neuron}, 56:\penalty0 530--540, November 2007.
\newblock ISSN 0896-6273.
\newblock \doi{10.1016/j.neuron.2007.08.017}.

\bibitem[Marks and Goard(2021)]{Marks2021}
Tyler~D. Marks and Michael~J. Goard.
\newblock Stimulus-dependent representational drift in primary visual cortex.
\newblock \emph{Nature Communications}, 12\penalty0 (1), aug 2021.
\newblock \doi{10.1038/s41467-021-25436-3}.

\bibitem[Micou and O'Leary(2023)]{Micou2023}
Charles Micou and Timothy O'Leary.
\newblock Representational drift as a window into neural and behavioural plasticity.
\newblock \emph{Curr. Opin. Neurobiol.}, 81\penalty0 (102746):\penalty0 102746, August 2023.

\bibitem[Morales et~al.(2025)Morales, Muñoz, and Tu]{Morales2025}
Guillermo~B. Morales, Miguel~A. Muñoz, and Yuhai Tu.
\newblock Representational drift and learning-induced stabilization in the piriform cortex.
\newblock \emph{Proceedings of the National Academy of Sciences}, 122\penalty0 (29):\penalty0 e2501811122, 2025.
\newblock \doi{10.1073/pnas.2501811122}.
\newblock URL \url{https://www.pnas.org/doi/abs/10.1073/pnas.2501811122}.

\bibitem[Natrajan and Fitzgerald(2025)]{Natrajan2025}
Maanasa Natrajan and James~E Fitzgerald.
\newblock Stability through plasticity: Finding robust memories through representational drift.
\newblock \emph{Proc. Natl. Acad. Sci. U. S. A.}, 122\penalty0 (45):\penalty0 e2500077122, November 2025.

\bibitem[Olshausen and Field(1996)]{Olshausen1996}
B~A Olshausen and D~J Field.
\newblock Natural image statistics and efficient coding.
\newblock \emph{Network}, 7\penalty0 (2):\penalty0 333--339, May 1996.

\bibitem[Papyan et~al.(2020{\natexlab{a}})Papyan, Han, and Donoho]{Papyan2020}
Vardan Papyan, X.~Y. Han, and David~L. Donoho.
\newblock Prevalence of neural collapse during the terminal phase of deep learning training.
\newblock \emph{Proceedings of the National Academy of Sciences of the United States of America}, 117:\penalty0 24652--24663, October 2020{\natexlab{a}}.
\newblock ISSN 1091-6490.
\newblock \doi{10.1073/pnas.2015509117}.

\bibitem[Papyan et~al.(2020{\natexlab{b}})Papyan, Han, and Donoho]{Papyan_2020}
Vardan Papyan, X.~Y. Han, and David~L. Donoho.
\newblock Prevalence of neural collapse during the terminal phase of deep learning training.
\newblock \emph{Proceedings of the National Academy of Sciences}, 117\penalty0 (40):\penalty0 24652--24663, sep 2020{\natexlab{b}}.
\newblock \doi{10.1073/pnas.2015509117}.

\bibitem[Pashakhanloo and Koulakov(2023)]{Pashakanloo2023}
Farhad Pashakhanloo and Alexei Koulakov.
\newblock Stochastic gradient descent-induced drift of representation in a two-layer neural network.
\newblock In \emph{Proceedings of the 40th International Conference on Machine Learning}, ICML'23. JMLR.org, 2023.

\bibitem[Pedregosa et~al.(2011{\natexlab{a}})Pedregosa, Varoquaux, Gramfort, Michel, Thirion, Grisel, Blondel, Prettenhofer, Weiss, Dubourg, Vanderplas, Passos, Cournapeau, Brucher, Perrot, and Duchesnay]{scikit-learn}
F.~Pedregosa, G.~Varoquaux, A.~Gramfort, V.~Michel, B.~Thirion, O.~Grisel, M.~Blondel, P.~Prettenhofer, R.~Weiss, V.~Dubourg, J.~Vanderplas, A.~Passos, D.~Cournapeau, M.~Brucher, M.~Perrot, and E.~Duchesnay.
\newblock Scikit-learn: Machine learning in {P}ython.
\newblock \emph{Journal of Machine Learning Research}, 12:\penalty0 2825--2830, 2011{\natexlab{a}}.

\bibitem[Pedregosa et~al.(2011{\natexlab{b}})Pedregosa, Varoquaux, Gramfort, Michel, Thirion, Grisel, Blondel, Prettenhofer, Weiss, Dubourg, Vanderplas, Passos, Cournapeau, Brucher, Perrot, and {{\'E}}douard Duchesnay]{JMLR:v12:pedregosa11a}
Fabian Pedregosa, Ga{{\"e}}l Varoquaux, Alexandre Gramfort, Vincent Michel, Bertrand Thirion, Olivier Grisel, Mathieu Blondel, Peter Prettenhofer, Ron Weiss, Vincent Dubourg, Jake Vanderplas, Alexandre Passos, David Cournapeau, Matthieu Brucher, Matthieu Perrot, and {{\'E}}douard Duchesnay.
\newblock Scikit-learn: Machine learning in python.
\newblock \emph{Journal of Machine Learning Research}, 12\penalty0 (85):\penalty0 2825--2830, 2011{\natexlab{b}}.
\newblock URL \url{http://jmlr.org/papers/v12/pedregosa11a.html}.

\bibitem[Qin et~al.(2023)Qin, Farashahi, Lipshutz, Sengupta, Chklovskii, and Pehlevan]{Qin2023}
Shanshan Qin, Shiva Farashahi, David Lipshutz, Anirvan~M Sengupta, Dmitri~B Chklovskii, and Cengiz Pehlevan.
\newblock Coordinated drift of receptive fields in {Hebbian/anti-Hebbian} network models during noisy representation learning.
\newblock \emph{Nat. Neurosci.}, 26\penalty0 (2):\penalty0 339--349, February 2023.

\bibitem[Radford et~al.(2021)Radford, Kim, Hallacy, Ramesh, Goh, Agarwal, Sastry, Askell, Mishkin, Clark, Krueger, and Sutskever]{CLIP2021}
Alec Radford, Jong~Wook Kim, Chris Hallacy, Aditya Ramesh, Gabriel Goh, Sandhini Agarwal, Girish Sastry, Amanda Askell, Pamela Mishkin, Jack Clark, Gretchen Krueger, and Ilya Sutskever.
\newblock Learning transferable visual models from natural language supervision.
\newblock In Marina Meila and Tong Zhang, editors, \emph{Proceedings of the 38th International Conference on Machine Learning}, volume 139 of \emph{Proceedings of Machine Learning Research}, pages 8748--8763. PMLR, 18--24 Jul 2021.
\newblock URL \url{https://proceedings.mlr.press/v139/radford21a.html}.

\bibitem[Resulaj et~al.(2018)Resulaj, Ruediger, Olsen, and Scanziani]{Resulaj2018}
Arbora Resulaj, Sarah Ruediger, Shawn~R Olsen, and Massimo Scanziani.
\newblock First spikes in visual cortex enable perceptual discrimination.
\newblock \emph{eLife}, 7, April 2018.
\newblock ISSN 2050-084X.
\newblock \doi{10.7554/elife.34044}.

\bibitem[Ruan et~al.(2022)Ruan, Dubois, and Maddison]{Ruan2021}
Yangjun Ruan, Yann Dubois, and Chris~J. Maddison.
\newblock Optimal representations for covariate shift.
\newblock In \emph{International Conference on Learning Representations}, 2022.
\newblock URL \url{https://openreview.net/forum?id=Rf58LPCwJj0}.

\bibitem[Ruderman and Bialek(1994)]{Ruderman1994}
D~L Ruderman and W~Bialek.
\newblock Statistics of natural images: Scaling in the woods.
\newblock \emph{Phys. Rev. Lett.}, 73\penalty0 (6):\penalty0 814--817, August 1994.

\bibitem[Rule and O'Leary(2022)]{Rule2022}
Michael~E. Rule and Timothy O'Leary.
\newblock Self-healing codes: How stable neural populations can track continually reconfiguring neural representations.
\newblock \emph{Proceedings of the National Academy of Sciences of the United States of America}, 119, February 2022.
\newblock ISSN 1091-6490.
\newblock \doi{10.1073/pnas.2106692119}.

\bibitem[Rule et~al.(2019)Rule, O'Leary, and Harvey]{Rule2019}
Michael~E. Rule, Timothy O'Leary, and Christopher~D. Harvey.
\newblock Causes and consequences of representational drift.
\newblock \emph{Current opinion in neurobiology}, 58:\penalty0 141--147, October 2019.
\newblock ISSN 1873-6882.
\newblock \doi{10.1016/j.conb.2019.08.005}.

\bibitem[Rule et~al.(2020)Rule, Loback, Raman, Driscoll, Harvey, and O'Leary]{Rule2020}
Michael~E. Rule, Adrianna~R. Loback, Dhruva~V. Raman, Laura~N. Driscoll, Christopher~D. Harvey, and Timothy O'Leary.
\newblock Stable task information from an unstable neural population.
\newblock \emph{eLife}, 9, July 2020.
\newblock ISSN 2050-084X.
\newblock \doi{10.7554/eLife.51121}.

\bibitem[Sadeh and Clopath(2022)]{Sadeh2022}
Sadra Sadeh and Claudia Clopath.
\newblock Contribution of behavioural variability to representational drift.
\newblock \emph{eLife}, 11:\penalty0 e77907, aug 2022.
\newblock ISSN 2050-084X.
\newblock \doi{10.7554/eLife.77907}.
\newblock URL \url{https://doi.org/10.7554/eLife.77907}.

\bibitem[Safran and Shamir(2018)]{Safran2017}
Itay Safran and Ohad Shamir.
\newblock Spurious local minima are common in two-layer {R}e{LU} neural networks.
\newblock In Jennifer Dy and Andreas Krause, editors, \emph{Proceedings of the 35th International Conference on Machine Learning}, volume~80 of \emph{Proceedings of Machine Learning Research}, pages 4433--4441. PMLR, 10--15 Jul 2018.
\newblock URL \url{https://proceedings.mlr.press/v80/safran18a.html}.

\bibitem[Salisbury and Palmer(2025)]{Salisbury2025}
Jared~M Salisbury and Stephanie~E Palmer.
\newblock A dynamic scale-mixture model of motion in natural scenes.
\newblock \emph{Elife}, 14\penalty0 (e104054), December 2025.

\bibitem[Saremi and Sejnowski(2013)]{Saremi2013}
Saeed Saremi and Terrence~J. Sejnowski.
\newblock Hierarchical model of natural images and the origin of scale invariance.
\newblock \emph{Proceedings of the National Academy of Sciences}, 110\penalty0 (8):\penalty0 3071--3076, February 2013.
\newblock ISSN 1091-6490.
\newblock \doi{10.1073/pnas.1222618110}.

\bibitem[Savier et~al.(2019)Savier, Chen, and Cang]{Savier2019}
Elise~L Savier, Hui Chen, and Jianhua Cang.
\newblock Effects of locomotion on visual responses in the mouse superior colliculus.
\newblock \emph{J. Neurosci.}, 39\penalty0 (47):\penalty0 9360--9368, November 2019.

\bibitem[Schneider et~al.(2023)Schneider, Lee, and Mathis]{schneider2023cebra}
Steffen Schneider, Jin~Hwa Lee, and Mackenzie~Weygandt Mathis.
\newblock Learnable latent embeddings for joint behavioural and neural analysis.
\newblock \emph{Nature}, May 2023.
\newblock ISSN 1476-4687.
\newblock \doi{10.1038/s41586-023-06031-6}.
\newblock URL \url{https://doi.org/10.1038/s41586-023-06031-6}.

\bibitem[Schoonover et~al.(2021)Schoonover, Ohashi, Axel, and Fink]{Schoonover2021}
Carl~E Schoonover, Sarah~N Ohashi, Richard Axel, and Andrew J~P Fink.
\newblock Representational drift in primary olfactory cortex.
\newblock \emph{Nature}, 594\penalty0 (7864):\penalty0 541--546, June 2021.

\bibitem[Schwartz and Simoncelli(2001)]{Schwartz2001}
O~Schwartz and E~P Simoncelli.
\newblock Natural signal statistics and sensory gain control.
\newblock \emph{Nat. Neurosci.}, 4\penalty0 (8):\penalty0 819--825, August 2001.

\bibitem[Shahinfar et~al.(2020)Shahinfar, Meek, and Falzon]{shahinfar2020many}
Saleh Shahinfar, Paul Meek, and Gregory Falzon.
\newblock How many images do i need? understanding how sample size per class affects deep learning model performance metrics for balanced designs in autonomous wildlife monitoring.
\newblock \emph{Ecological Informatics}, 57:\penalty0 101085, 2020.

\bibitem[Siegle et~al.(2021)Siegle, Jia, Durand, Gale, Bennett, Graddis, Heller, Ramirez, Choi, Luviano, Groblewski, Ahmed, Arkhipov, Bernard, Billeh, Brown, Buice, Cain, Caldejon, Casal, Cho, Chvilicek, Cox, Dai, Denman, de~Vries, Dietzman, Esposito, Farrell, Feng, Galbraith, Garrett, Gelfand, Hancock, Harris, Howard, Hu, Hytnen, Iyer, Jessett, Johnson, Kato, Kiggins, Lambert, Lecoq, Ledochowitsch, Lee, Leon, Li, Liang, Long, Mace, Melchior, Millman, Mollenkopf, Nayan, Ng, Ngo, Nguyen, Nicovich, North, Ocker, Ollerenshaw, Oliver, Pachitariu, Perkins, Reding, Reid, Robertson, Ronellenfitch, Seid, Slaughterbeck, Stoecklin, Sullivan, Sutton, Swapp, Thompson, Turner, Wakeman, Whitesell, Williams, Williford, Young, Zeng, Naylor, Phillips, Reid, Mihalas, Olsen, and Koch]{Siegle2021}
Joshua~H. Siegle, Xiaoxuan Jia, S{\'{e}}verine Durand, Sam Gale, Corbett Bennett, Nile Graddis, Greggory Heller, Tamina~K. Ramirez, Hannah Choi, Jennifer~A. Luviano, Peter~A. Groblewski, Ruweida Ahmed, Anton Arkhipov, Amy Bernard, Yazan~N. Billeh, Dillan Brown, Michael~A. Buice, Nicolas Cain, Shiella Caldejon, Linzy Casal, Andrew Cho, Maggie Chvilicek, Timothy~C. Cox, Kael Dai, Daniel~J. Denman, Saskia E.~J. de~Vries, Roald Dietzman, Luke Esposito, Colin Farrell, David Feng, John Galbraith, Marina Garrett, Emily~C. Gelfand, Nicole Hancock, Julie~A. Harris, Robert Howard, Brian Hu, Ross Hytnen, Ramakrishnan Iyer, Erika Jessett, Katelyn Johnson, India Kato, Justin Kiggins, Sophie Lambert, Jerome Lecoq, Peter Ledochowitsch, Jung~Hoon Lee, Arielle Leon, Yang Li, Elizabeth Liang, Fuhui Long, Kyla Mace, Jose Melchior, Daniel Millman, Tyler Mollenkopf, Chelsea Nayan, Lydia Ng, Kiet Ngo, Thuyahn Nguyen, Philip~R. Nicovich, Kat North, Gabriel~Koch Ocker, Doug Ollerenshaw, Michael Oliver, Marius Pachitariu, Jed
  Perkins, Melissa Reding, David Reid, Miranda Robertson, Kara Ronellenfitch, Sam Seid, Cliff Slaughterbeck, Michelle Stoecklin, David Sullivan, Ben Sutton, Jackie Swapp, Carol Thompson, Kristen Turner, Wayne Wakeman, Jennifer~D. Whitesell, Derric Williams, Ali Williford, Rob Young, Hongkui Zeng, Sarah Naylor, John~W. Phillips, R.~Clay Reid, Stefan Mihalas, Shawn~R. Olsen, and Christof Koch.
\newblock Survey of spiking in the mouse visual system reveals functional hierarchy.
\newblock \emph{Nature}, 592\penalty0 (7852):\penalty0 86--92, jan 2021.
\newblock \doi{10.1038/s41586-020-03171-x}.

\bibitem[Sj{\"o}str{\"o}m et~al.(2004)Sj{\"o}str{\"o}m, Turrigiano, and Nelson]{Sjostrom2004}
Per~Jesper Sj{\"o}str{\"o}m, Gina~G Turrigiano, and Sacha~B Nelson.
\newblock Endocannabinoid-dependent neocortical layer-5 {LTD} in the absence of postsynaptic spiking.
\newblock \emph{J. Neurophysiol.}, 92\penalty0 (6):\penalty0 3338--3343, December 2004.

\bibitem[Stringer et~al.(2019{\natexlab{a}})Stringer, Pachitariu, Steinmetz, Carandini, and Harris]{Stringer2019a}
Carsen Stringer, Marius Pachitariu, Nicholas Steinmetz, Matteo Carandini, and Kenneth~D. Harris.
\newblock High-dimensional geometry of population responses in visual cortex.
\newblock \emph{Nature}, 571:\penalty0 361--365, July 2019{\natexlab{a}}.
\newblock ISSN 1476-4687.
\newblock \doi{10.1038/s41586-019-1346-5}.

\bibitem[Stringer et~al.(2019{\natexlab{b}})Stringer, Pachitariu, Steinmetz, Reddy, Carandini, and Harris]{Stringer2019}
Carsen Stringer, Marius Pachitariu, Nicholas Steinmetz, Charu~Bai Reddy, Matteo Carandini, and Kenneth~D. Harris.
\newblock Spontaneous behaviors drive multidimensional, brainwide activity.
\newblock \emph{Science}, 364\penalty0 (6437), apr 2019{\natexlab{b}}.
\newblock \doi{10.1126/science.aav7893}.

\bibitem[Tian et~al.(2019)Tian, Krishnan, and Isola]{Tian2019}
Yonglong Tian, Dilip Krishnan, and Phillip Isola.
\newblock Contrastive multiview coding.
\newblock June 2019.

\bibitem[Tian et~al.(2020)Tian, Krishnan, and Isola]{tian2020contrastive}
Yonglong Tian, Dilip Krishnan, and Phillip Isola.
\newblock Contrastive multiview coding.
\newblock In \emph{European Conference on Computer Vision}, pages 776--794. Springer, 2020.

\bibitem[Turrigiano et~al.(1998)Turrigiano, Leslie, Desai, Rutherford, and Nelson]{Turrigiano1998}
Gina~G Turrigiano, Kenneth~R Leslie, Niraj~S Desai, Lana~C Rutherford, and Sacha~B Nelson.
\newblock Activity-dependent scaling of quantal amplitude in neocortical neurons.
\newblock \emph{Nature}, 391\penalty0 (6670):\penalty0 892--896, February 1998.

\bibitem[van Versendaal and Levelt(2016)]{Van_Versendaal2016}
Dani{\"e}lle van Versendaal and Christiaan~N Levelt.
\newblock Inhibitory interneurons in visual cortical plasticity.
\newblock \emph{Cell. Mol. Life Sci.}, 73\penalty0 (19):\penalty0 3677--3691, October 2016.

\bibitem[Wainwright and Simoncelli(1999)]{Wainwright2000}
Martin~J Wainwright and Eero Simoncelli.
\newblock Scale mixtures of gaussians and the statistics of natural images.
\newblock In S.~Solla, T.~Leen, and K.~M\"{u}ller, editors, \emph{Advances in Neural Information Processing Systems}, volume~12. MIT Press, 1999.
\newblock URL \url{https://proceedings.neurips.cc/paper_files/paper/1999/file/6a5dfac4be1502501489fc0f5a24b667-Paper.pdf}.

\bibitem[Wainwright et~al.(2001)Wainwright, Simoncelli, and Willsky]{Wainwright2001}
Martin~J Wainwright, Eero~P Simoncelli, and Alan~S Willsky.
\newblock Random cascades on wavelet trees and their use in analyzing and modeling natural images.
\newblock \emph{Appl. Comput. Harmon. Anal.}, 11\penalty0 (1):\penalty0 89--123, July 2001.

\bibitem[Wang et~al.(2022)Wang, Hoshal, de~Laittre, Marre, Berry, and Palmer]{wang2022learning}
Siwei Wang, Benjamin Hoshal, Elizabeth~A de~Laittre, Olivier Marre, Michael Berry, and Stephanie Palmer.
\newblock Learning low-dimensional generalizable natural features from retina using a u-net.
\newblock In Alice~H. Oh, Alekh Agarwal, Danielle Belgrave, and Kyunghyun Cho, editors, \emph{Advances in Neural Information Processing Systems}, 2022.
\newblock URL \url{https://openreview.net/forum?id=bdfJCeWDKUB}.

\bibitem[Wang and Oates(2015)]{Wang2015}
Zhiguang Wang and Tim Oates.
\newblock Imaging time-series to improve classification and imputation.
\newblock In \emph{Proceedings of the 24th International Conference on Artificial Intelligence}, IJCAI'15, page 3939–3945, Buenos Aires, Argentina, 2015. AAAI Press.
\newblock ISBN 9781577357384.

\bibitem[Ward(1963)]{Ward63}
Joe~H. Ward.
\newblock Hierarchical grouping to optimize an objective function.
\newblock \emph{Journal of the American Statistical Association}, 58\penalty0 (301):\penalty0 236--244, 1963.
\newblock URL \url{http://www.jstor.org/stable/2282967}.

\bibitem[Wen and Turrigiano(2024)]{Wen2024}
Wei Wen and Gina~G Turrigiano.
\newblock Keeping your brain in balance: Homeostatic regulation of network function.
\newblock \emph{Annu. Rev. Neurosci.}, 47\penalty0 (1):\penalty0 41--61, August 2024.

\bibitem[Xia et~al.(2021)Xia, Marks, Goard, and Wessel]{Xia2021}
Ji~Xia, Tyler~D. Marks, Michael~J. Goard, and Ralf Wessel.
\newblock Stable representation of a naturalistic movie emerges from episodic activity with gain variability.
\newblock \emph{Nature communications}, 12:\penalty0 5170, August 2021.
\newblock ISSN 2041-1723.
\newblock \doi{10.1038/s41467-021-25437-2}.

\bibitem[Xu and Raginsky(2020)]{Xu2020}
Aolin Xu and Maxim Raginsky.
\newblock Minimum excess risk in bayesian learning.
\newblock December 2020.

\bibitem[Yilmaz and Meister(2013)]{Yilmaz2013}
Melis Yilmaz and Markus Meister.
\newblock Rapid innate defensive responses of mice to looming visual stimuli.
\newblock \emph{Curr. Biol.}, 23\penalty0 (20):\penalty0 2011--2015, October 2013.

\bibitem[Yoon and Joo(2025)]{Yoon2025}
Gi-Won Yoon and Segyeong Joo.
\newblock Enhanced electrocardiogram classification using gramian angular field transformation with multi-lead analysis and segmentation techniques.
\newblock \emph{MethodsX}, 14:\penalty0 103297, June 2025.
\newblock ISSN 2215-0161.
\newblock \doi{10.1016/j.mex.2025.103297}.
\newblock URL \url{http://dx.doi.org/10.1016/j.mex.2025.103297}.

\bibitem[Yun et~al.(2017)Yun, Sra, and Jadbabaie]{Yun2017}
Chulhee Yun, Suvrit Sra, and Ali Jadbabaie.
\newblock Global optimality conditions for deep neural networks.
\newblock \emph{CoRR}, abs/1707.02444, 2017.
\newblock URL \url{http://arxiv.org/abs/1707.02444}.

\bibitem[Yun et~al.(2019)Yun, Sra, and Jadbabaie]{Yun2018}
Chulhee Yun, Suvrit Sra, and Ali Jadbabaie.
\newblock Small nonlinearities in activation functions create bad local minima in neural networks.
\newblock In \emph{International Conference on Learning Representations}, 2019.
\newblock URL \url{https://openreview.net/forum?id=rke_YiRct7}.

\bibitem[Zhang et~al.(2014)Zhang, Xu, Kamigaki, Hoang~Do, Chang, Jenvay, Miyamichi, Luo, and Dan]{Zhang2014}
Siyu Zhang, Min Xu, Tsukasa Kamigaki, Johnny~Phong Hoang~Do, Wei-Cheng Chang, Sean Jenvay, Kazunari Miyamichi, Liqun Luo, and Yang Dan.
\newblock Selective attention. long-range and local circuits for top-down modulation of visual cortex processing.
\newblock \emph{Science (New York, N.Y.)}, 345:\penalty0 660--665, August 2014.
\newblock ISSN 1095-9203.
\newblock \doi{10.1126/science.1254126}.

\bibitem[Zhao et~al.(2014)Zhao, Liu, and Cang]{Zhao2014}
Xinyu Zhao, Mingna Liu, and Jianhua Cang.
\newblock Visual cortex modulates the magnitude but not the selectivity of looming-evoked responses in the superior colliculus of awake mice.
\newblock \emph{Neuron}, 84\penalty0 (1):\penalty0 202--213, October 2014.

\bibitem[Zhou et~al.(2021)Zhou, Ge, and Jin]{Zhou2021}
Mo~Zhou, Rong Ge, and Chi Jin.
\newblock A local convergence theory for mildly over-parameterized two-layer neural network.
\newblock \emph{CoRR}, abs/2102.02410, 2021.
\newblock URL \url{https://arxiv.org/abs/2102.02410}.

\bibitem[Zhu et~al.(2018)Zhu, Soudry, Eldar, and Wakin]{Zhu2018}
Zhihui Zhu, Daniel Soudry, Yonina~C. Eldar, and Michael~B. Wakin.
\newblock The global optimization geometry of shallow linear neural networks.
\newblock \emph{CoRR}, abs/1805.04938, 2018.
\newblock URL \url{http://arxiv.org/abs/1805.04938}.

\bibitem[Zhu et~al.(2021)Zhu, Ding, Zhou, Li, You, Sulam, and Qu]{Zhu2021}
Zhihui Zhu, Tianyu Ding, Jinxin Zhou, Xiao Li, Chong You, Jeremias Sulam, and Qing Qu.
\newblock A geometric analysis of neural collapse with unconstrained features.
\newblock In \emph{Proceedings of the 35th International Conference on Neural Information Processing Systems}, NIPS '21, Red Hook, NY, USA, 2021. Curran Associates Inc.
\newblock ISBN 9781713845393.

\end{thebibliography}
\end{document}